\documentclass{aa}
\usepackage{graphicx}
\usepackage{txfonts}
\usepackage{epsfig,graphics,graphicx,bm,amssymb}
\begin{document}

\title{Formation and evolution of the two 4/3 resonant giants planets in HD\,200946}

\author{M. Tadeu dos Santos\thanks{e-mail: mtadeu@astro.iag.usp.br}
          \and \, J. A. Correa-Otto\thanks{e-mail: jorge9895@usp.br}
          \and \,\, T. A. Michtchenko
          \and S. Ferraz-Mello}

\institute{Instituto de Astronomia, Geof\'isica e Ci\^encias Atmosf\'ericas, USP, Rua do Mat\~ao 1226, 05508-090 S\~ao
Paulo, Brazil}
 
   \date{}

\abstract{It has been suggested that HD\,200964 is the first exoplanetary system with two Jovian planets evolving in the 4/3 mean-motion resonance. Previous scenarios to simulate the formation of two giant planets in the stable 4/3 resonance configuration have failed. Moreover, the orbital parameters available in the literature point out an unstable configuration of the planetary pair.}
{
The purpose of this paper is i) to determine the orbits of the planets from the RV measurements and update the value of the stellar mass ($1.57\,M_{\odot}$),  ii) to analyse the stability of the planetary evolution in the vicinity and inside the 4/3 MMR, and iii) to elaborate a possible scenario for the formation of systems in the 4/3 MMR.
}
{
We use a previous model to simulate the formation of the stable planetary pair trapped inside the 4/3 resonance. Our scenario includes an interaction between the type I and type II of migration, planetary growth and stellar evolution from the main sequence to the sub-giant branch. The redetermination of the orbits is done using a biased Monte Carlo procedure, while the planetary dynamics is studied using numerical tools, such as dynamical maps and dynamical power spectra.
}
{
The results of the formation simulations are able to very closely reproduce the 4/3 resonant dynamics of the best-fit configuration obtained in this paper. Moreover, the confidence interval of the fit matches well  with the very narrow stable region of the 4/3 mean-motion resonance.
}
{
The formation process of the HD\,200964 system is very sensitive to the planetary masses and protoplanetary disk parameters. Only a thin, flat disk allows the embryo-sized planets to reach the 4/3 resonant configuration. The stable evolution of the resonant planets is also sensitive to the mass of the central star, because of overlapping high-order resonances inside the 4/3 resonance. Regardless of the very narrow domain of stable motion, the confidence interval of our fit closely matches the stability area.
}


\titlerunning{Formation and evolution of the HD\,200964 system}

\maketitle

\section{Introduction}\label{sect-0}

HD\,200964 is the first discovered system with two giant planets to evolve inside the 4/3 mean-motion resonance (Johnson et al. 2011). The planets revolve around a sub-giant K-type star with estimated lifetime  3\,Gyr and mass  $1.57\,M_{\odot}$, according to the recent estimate given by Mortier et al. (2013). The determination of the planetary orbits was done using the 61 Lick and 35 Keck radial velocity measurements (RV) and the dynamical behavior of the system in the domain around the nominal best-fit solution was analyzed over 10\,Myr by Johnson et al. (2011). The results have shown that the current observations strongly favor orbital solutions in or close to the 4/3 mean-motion resonance (MMR).

Wittenmyer et al. (2012) have extended  over 100\,Myr the analysis of stability of the HD\,200964 system mapping the neighborhood of the solution proposed by Johnson et al. (2011). Their results have confirmed that the long-lived planetary pair of giants is only possible if the planets are currently trapped inside the mutual 4/3 MMR.  However, none of the 4/3 resonant configurations tested in their simulations has survived over 100\,Myr. Regardless of the very long extension of the tested timespan, it is still one order less than the lifetime of the central star. This raises a question of whether a system composed of such close Jupiter-sized planets, evolving on mutually crossing orbits, is reliable.

The origin of the HD\,200964 system is another unsolved problem. The issue was extensively explored in Rein et al. (2012), where the model of formation of planetary pairs in the 4/3 MMR based on the traditional type II migration scenario was investigated. However, even employing non-conventional scenarios, such as formation in situ and planetary scattering,  these attempts to simulate the formation of two giant planets in the stable 4/3 resonance configuration have failed.

It is clear that the inherent problems described above  require a new approach in the study of formation and evolution of such close and massive planetary systems like the HD\,200964 system. In this paper, we introduce the approach whose principal components are i) the redetermination of the orbits of the planets in the HD\,200964 system from the RV measurements, ii) the complete understanding of the planetary evolution in the vicinity of and inside the 4/3 MMR, and iii) the elaboration of a possible scenario of the formation of systems in the 4/3 MMR. We show that only the joint consideration of all three factors can allow us to simulate a reliable HD\,200964 planetary system.

We use various models and methods developed and described in  previous papers. For instance, the scenario of formation of close planetary systems formed of two giants has been developed in Correa-Otto et al. (2013). It includes an interaction between different planetary migration types (types I and II), planetary growth, and gap formation in the protoplanetary disk. The joint action of these mechanisms has proved to be efficient at reproducing the  behavior of the HD\,45364 system with two close and massive planets currently evolving in the 3/2 MMR (Correia et al. 2009). It is worth noting here that the choice of this formation model in this work was done after the detailed investigation of the planetary dynamics in the neighborhood of the 4/3 MMR. The study was done in the form of dynamical maps (Michtchenko et al. 2002, Ferraz-Mello et al. 2005), which showed that the resonance capture would be possible only in the stage when the planets were still embryo-sized.

The planetary configuration of the solution obtained in the simulation of formation is then tested in order to verify if the system could survive over the whole lifetime of the central star.  This is frequently done by means of purely numerical integrations of the exact equations  of motion over extended timespans (e.g., Wittenmyer et al. 2012). In this work, we apply an alternative approach which focuses on the general dynamics of systems involved in mean-motion resonances (Beaug\'e et al. 2003, Ferraz-Mello et al. 2006, Michtchenko et al. 2006, among others). In practice, instead of testing the stability of individual solutions, we determine the domains of stable motion near the 4/3 MMR (for details, see Michtchenko et al. 2008\,a,b). These domains are located around stable stationary configurations (known as apsidal corotation resonances, ACRs), which are obtained applying the Hamiltonian model of the resonant behavior to the 4/3 MMR (Michtchenko et al. 2006). The stable ACRs constrain the planetary geometry favorable for the regular evolution in the 4/3 MMR when the orbits are anti-aligned; at the conjunction, the inner planet is at the pericenter of its orbit, while the outer planet is at apocenter.

The stable domains of the 4/3 MMR are delimited by numerous overlapping high-order mean-motion resonances (Wisdom 1980) whose locations and widths are very sensitive  to the changes in the stellar and planetary masses. This feature allows us to put constraints on the masses and the inclination of the HD\,200964 orbits to the sky plane.

The dynamical stability of the HD\,200964 system is also tested considering the variation of the mass of the central star during the stellar evolution from the main sequence (MS) to the sub-giant branch.  For this, we estimate the main sequence mass of the HD\,200964 star using the empirical relationship between the age of the star and its mass at the MS stage (Inglis 2003), and  obtain that the mass loss was $\sim$ 3\,\%. We consider the possible implications of the stellar mass loss for planetary dynamics and show that, for small values, the stability of the planetary motion is not affected.  This result is in agreement with the results obtained by Voyatzis et al. (2013), which showed that the planetary motion is destabilized  when the loss of the stellar mass reaches $10\,\%$.

Finally, we redetermine the orbits of the planets from the radial velocity measurements given in Jonhson et al. (2011), where the best-fit solution was obtained  using the non-updated stellar mass $1.44\,M_{\odot}$. Our main goal was to extend the search, looking for the possible solutions out of the 4/3 MMR.  The analysis was based on the biased Monte Carlo procedure (Ferraz-Mello et al. 2005, Tadeu dos Santos et al. 2012), that uses a standard orbit improvement technique  with starting values taken at random in a large set of initial conditions. The constraints on the planetary motions provided by the previous stability study were also implemented. As a result, we obtained the confidence interval of statistically good and stable solutions inside the 4/3 MMR, compatible with the solution obtained from the simulation of formation.

The paper is organized as follows. In Sect. \ref{sect-1}, we present dynamical maps of the region between the 2/1 and 4/3 MMRs. In Sect. \ref{sect-2}, we explore the formation scenario, including stellar mass evolution, and emphasize the conditions necessary for the formation of the HD\,200964 system. In Sect. \ref{sect-3}, we develop a dynamical study of the 4/3 MMR, employing the dynamical maps and dynamical power spectra techniques. In Sect. \ref{sect-4} we reanalyze the radial velocities data linking the orbital solutions obtained with the results described in Sect. \ref{sect-2}. Finally, in Sect. \ref{sect-5}, we summarize our results.

\section{Dynamical map of the region between the 2/1 and 4/3 resonances}\label{sect-1}

According to classical theories of migration, gravitational interactions between the protoplanetary disk and planets drive the planets towards the central star (Lin $\&$ Papaloizou 1979; Goldreich $\&$ Tremaine 1979, 1980). If the migration of a planetary pair is convergent (i.e., the mutual planetary distance is decreasing), the planets can be captured into a mean-motion commensurability and continue to decay evolving in the resonance (e.g., Lee $\&$ Peale 2002; Ferraz-Mello et al. 2003; Beaug\'e et al. 2003; Beaug\'e et al. 2006). This mechanism is widely accepted to explain the existence of many extra-solar systems currently evolving inside mean-motion resonances.

\begin{figure}
\begin{center}
  \includegraphics[width=0.97\columnwidth]{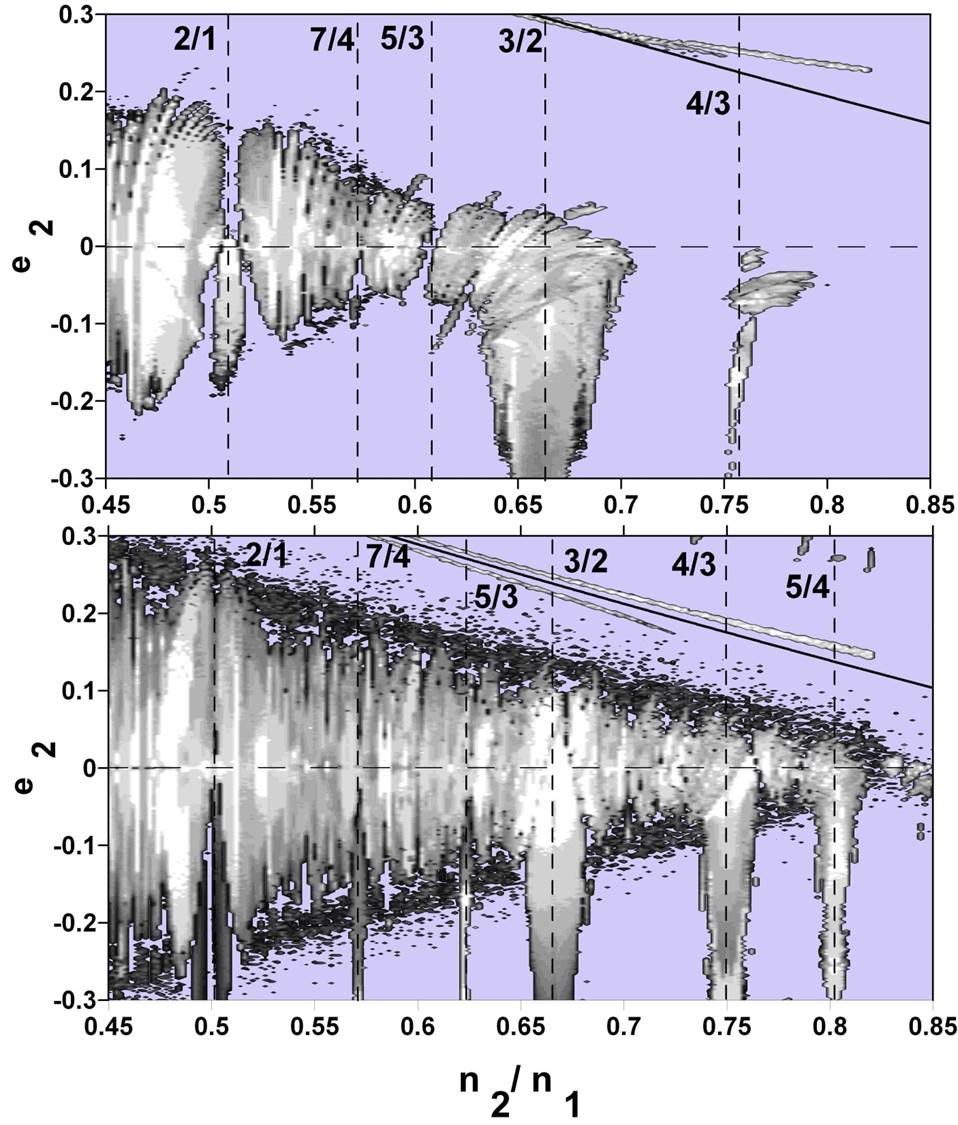}
\caption{Dynamical maps of the domain between the 2/1 and 4/3 MMRs, for two hypothetical systems composed of a solar mass star and two planets with masses $m_1 = m_2 = 1.0\,M_J$ (top panel) and $m_1 = m_2 = 0.1\,M_J$ (bottom panel). The   secular angle $\Delta \varpi$ is fixed at 0 (positive values on the $e_2$-axis) or $180^o$ (negative values on the $e_2$-axis).  The light grey tones correspond to regions of regular motion, while the regions in blue correspond to highly chaotic motion. The location of several MMRs is indicated by vertical dashed lines. The solid lines mark the boundaries of the domains of mutually crossing planetary orbits.}
\label{fig1}
\end{center}
\end{figure}

To obtain the possible migration routes of the system towards the 4/3 MMR, we analyze the dynamics of the planets in the region between the 2/1 and 4/3 MMRs. Using typical values of the physical and orbital parameters of exoplanets, we construct dynamical maps on the $(n_2/n_1, e_2)$ representative plane, where $n_1$ and $n_2$ are the osculating mean motions of the inner and outer planets, respectively, and $ e_2$ is the osculating eccentricity of the outer orbit. The maps are presented in Fig. \ref{fig1}. In the construction of the maps, each $(n_2/n_1, e_2)$--plane was covered with a rectangular grid of initial conditions, with spacings $\Delta (n_2/ n_1) = 0.002$ and $\Delta e_2=0.002$. The semimajor axis and the eccentricity of the inner planet were fixed at $a_1 = 1$\,AU and $e_1 = 0.001$.

The initial values of the mean longitudes were fixed at $\lambda_1=\lambda_2=0$, while the difference of the longitudes of the pericenter of the planets,  $\Delta \varpi = \varpi_2 - \varpi_1$, was fixed at either 0 (positive values on the $e_2$-axis in Figure \ref{fig1}) and $180^o$ (negative values on the $e_2$-axis). The chosen values of the angular elements correspond to the 0-- or $\pi$--values of the resonant angles of the mean-motion resonances which populate the domain under study. On the one hand, these configurations, also known as symmetric stationary solutions, are mostly favorable to stable resonant motions, especially at high eccentricities  (Ferraz-Mello et al. 2003, Michtchenko et al. 2006, Michtchenko et al. 2008\,a). On the other hand, from the point of view of capture in resonance, the angular elements can be chosen arbitrarily. Indeed, as will be shown in Section \ref{sect-2}, the probability of the resonance trapping is not affected by the choice of the angular elements, provided that capture occurs when the planetary orbits are nearly circular.

We used two different sets for the planetary masses: the map obtained for $m_1 = m_2 = 1.0\,M_J$ is shown in the top panel in Figure \ref{fig1}, while the map obtained for $m_1 = m_2 = 0.1\,M_J$ is shown in the bottom panel. The different values of the masses were used to understand qualitatively the dependence of the dynamical features on the individual masses of the planets.

Each point of the grid of initial conditions was numerically integrated over $1.3\,\times\,10^5$\,yr and the output of integration was Fourier analyzed in order to obtain the spectral number $N$  (see Spectral Analysis Method (SAM) in Michtchenko et al. 2002, Ferraz-Mello et al. 2005, Appendix). This quantity is defined as the number of peaks in the power spectrum of the eccentricity of the inner planet and is used to quantify the level of chaos in planetary motion: small values of $N$ correspond to regular motion, while the large values indicate the onset of chaos. The grey color scale is then used to plot the spectral number $N$ on the $(n_2/n_1, e_2)$ representative plane, where light regions correspond to the small spectral numbers $N$ and regular motions, while the increasingly dark tones indicate large  spectral numbers and unstable orbits. The blue-colored regions correspond to the domains of strongly chaotic motion leading to ejection/collision of the planets.

The map obtained for $m_1 = m_2 = 1.0\,M_J$ is shown on the top panel in Figure \ref{fig1}. The domains of the first-order 2/1, 3/2, and 4/3 MMRs are located around $n_2/n_1 = 0.509$, $0.663$, and $0.757$, respectively; each resonance is bounded by the chaotic layers, which correspond to the separatrix of the resonance. The robust structure of the 2/1 and 3/2 MMRs indicates that the smooth passages of the system during converging type II migration will probably be interrupted, either by capture inside the resonance (at smaller eccentricities) or by ejection of the planets (at higher eccentricities). 

There are other strong mean-motion resonances between 2/1 and 3/2 MMR: they are 7/4 and 5/3 commensurabilities located at $n_2/n_1\,\sim\,0.57$  and $0.61$, respectively. From these, the second-order 5/3 MMR is also sufficiently strong to capture the system during slow migration. It should be emphasized that resonant captures are fairly robust when a slow type II migration is considered (see Emel\'yanenko 2012, Correa-Otto et al. 2013, among others). Thus, the 2/1, 5/3, and 3/2 MMRs will act as natural dynamical barriers that prevent planets formed far apart to migrate slowly toward the 4/3 MMR.  Moreover, the large zone of strong instability separating the 4/3 MMR from the 3/2 MMR  makes the passage of the planetary pairs in the direction of the 4/3 MMR  impossible. Therefore, to explain the existence of the system HD\,200964 in the context of the type II migration, the two Jupiter-size planets are expected to be formed very close to each other, in a small region located practically on top of the 4/3 MMR. This hypothesis seems very unlikely, however.

In this work, we introduce an alternative scenario that describes the trapping in the 4/3 MMR of the embryo-sized planets that are still in the initial stage of the planetary formation. The main idea of our approach is clearly illustrated in the bottom panel in Figure \ref{fig1}, where we present a dynamical map analogous to that in the top panel, except that the individual planetary masses are ten times smaller. Since the mutual perturbations between the planets are weaker in this case, the width of the resonant domains  decreases (Michtchenko et al. 2008\,a,b); this is clearly seen in the case of the strong 2/1 and 3/2 MMRs. As a consequence, the non-capture passages through these resonances toward the 4/3 MMR become possible during the type I migration thanks to the small planetary masses. Moreover, the zone of strong instabilities between the 4/3 and 3/2 MMRs at low eccentricities disappears allowing the smooth crossing of the planets.


\section{Formation scenario of the HD\,200964 system}\label{sect-2}

The origin of the compact planetary configuration in the HD\,200964 system was first investigated by Rein et al. (2012). Various models for the formation of systems in the 4/3 MMR based on traditional type II migration scenarios or in situ formation of the planets were explored.  However, all attempts to simulate the formation of two massive planets in the stable 4/3 resonance configuration have failed.

In this work,  to explain the existence of the HD\,200964 system, we adopted the Alibert et al. (2005) scenario. We propose that capture of the planet pair in the 4/3 MMR occurred during the first phase of the planet growth, when the approximation of the system to the 4/3 MMR was favored by the small planetary masses and the fast type I migration. After the planets had been trapped in this resonance, they continued to grow to stage I, until their masses reached $\sim\,10\,M_{\oplus}$ (a typical critical mass value); then they evolved to stage II (runaway growth) and a slower type II migration. In our experiment, this process is separated schematically into two phases that are described below.

\begin{table*}

\centering
\caption{Physical and orbital parameters of the HD\,200964 system corresponding to the best fit of Johnson et al. (2011), our RV adjustment and the solution obtained from the simulation of planet formation (JD = 2453213.895 corresponding to the first observation from Jonhson et al. 2011). }\label{table1-2}

\begin{tabular}{l|l|ll|ll|l}
\hline\hline
Object &  Parameter  & \multicolumn{2}{|c|}{Johnson et al. (2011) } & \multicolumn{2}{|c|}{ RV adjustment } & Simulated \\
  &     & \multicolumn{2}{|c|}{  } & \multicolumn{2}{|c|}{ (without jitter) } & solution \\[0.7ex]
          &             &   median  &   confidence        &           median   &   confidence     &             \\
           &             &   values  &    interval       &            values  &   interval    &             \\
\hline
   HD\,200964  & Mass ($M_{\odot}$) &  $1.44 $   &   $-$        & $1.57 $  &   $-$                             &  $1.57$       \\
       &  &  &  & & & \\
   HD\,200964 b& $m\,{\rm sin}{I}$ ($M_J$)      &  $1.85 $ & $\left[1.73;2.0   \right]$   & $2.05$ &  $\left[1.5;2.2   \right]  $  &  $2.0$   \\
               & $a ({\rm AU})$           &  $1.60 $ & $\left[1.599;1.602   \right]$     & $1.68$ & $\left[1.6;1.7   \right]  $        &  $1.67$       \\
               & $e$               &  $0.040$ &   $\left[0.038;0.044   \right]$     & $0.085$ & $\left[0.01;0.15 \right]  $      &  $0.036$      \\
               & $\lambda$ (deg)     &  $ -   $ &   $-$          & $0.75$ & $\left[300;370   \right]  $       &  $101.6$        \\
               & $\varpi$ (deg)     &  $288  $ &  $\left[176;335   \right]$   & $184$ &  $\left[97;305    \right]  $              &  $151$        \\
       &  &  &  & & &  \\
   HD\,200964 c  & $m\,{\rm sin}{I} (M_J)$      &  $0.895 $ &  $\left[0.832;1.018   \right]$   & $0.89$  & $\left[0.85;1.45 \right]  $    &  $0.90$       \\
               & $a ({\rm AU})$           &  $1.95 $  &    $\left[1.945;1.958   \right]$        & $2.08$ & $\left[1.9;2.4   \right]  $   &  $2.03$       \\
               & $e$               &  $0.181$  &  $\left[0.164;0.205   \right]$         & $0.475$ & $\left[0.2;0.6   \right]  $   &  $0.122$      \\
               & $\lambda$ (deg)     &  $ -   $ & $-$     & $268$ &  $\left[260;370   \right]  $         &  $14.8$        \\
               & $\varpi$ (deg)     &  $182  $  &  $\left[125;250   \right]$   & $304$ &  $ \left[270;350   \right]  $     &  $293$        \\
\hline
  $w.r.m.s.$    &   $m/s$       &  \multicolumn{2}{|c|}{ $6.8$ }  & \multicolumn{2}{|c|}{ $5.26$  } & $7.8$  \\
\hline
 $\sqrt{\chi_\nu^2}$   &      & \multicolumn{2}{|c|}{1.15}  & \multicolumn{2}{|c|}{2.5 } & 3.7 \\
\hline
\end{tabular}
\end{table*}


\subsection{HD\,200964 star mass evaluation}\label{sect-2-2}

Masses of the stars are poorly determined quantities. Generally, this has no consequences in  dynamical stability studies, since the star mass uncertainties provoke no significant changes in the planet behavior (Beaug\'e $\&$ Michtchenko 2003). However, as will be shown in Sect. \ref{sect-3-2}, for close and massive planets (e.g., HD\,200964\,b-c), the dependence of the planetary dynamics on the mass of the central star is strong, with important implications for the stability and even physical integrity of the system. Thus, the mass evolution of the sub-giant HD\,200964 star must be taken into account in the investigations of planet formation. 

HD\,200964 is a sub-giant K0 IV star of age of $3\,\times\,10^9$\,years. Until recently, its current mass was estimated at 1.44\,$M_{\odot}$ and this value was used in the  determination of the planetary orbits in Jonhson et al. (2011). This estimate was recently revised and increased to 1.57\,$M_{\odot}$ by Mortier et al. (2013). According to astrophysical models, the lifetime of an evolved A star can be used to estimate the stellar mass during its stage in the MS (Inglis 2003); thus, the MS mass of the HD\,200964 star can be estimated as $\sim\,1.61\,M_{\odot}$  (see Table 3.1 in Inglis, 2003) and the mass loss during $3\,\times\,10^9$\,yr as $\sim\,0.04\,M_{\odot}$ ($3\,\%$). More details about the post-main-sequence evolution of the star will be presented in Sect. \ref{sect-2-6}.


\subsection{Stage I: core formation, type I migration and capture into the 4/3 MMR}\label{sect-2-3}

We assume that, in the first stage of the formation, the system is composed initially of the two planetary embryos with masses $m_1 = 0.1\,M_{\oplus}$ and $m_2 = 0.11\,M_{\oplus}$, orbiting the main sequence HD\,200964 star with mass $1.61\,M_{\odot}$. The planetary embryos are embedded in the protoplanetary disk, which we assume to be vertically isothermal and laminar. The interactions of the growing planets with the disk originate the type I convergent migration toward the central star. From the observational data, the lifetime of protoplanetary disks is estimated to lie between 1-10 Myr (Haisch et al. 2001). Also, some observational studies relate the mass of the disk to the mass of the central star introducing the median disk-to-star mass ratio of $\sim\,0.5\,\%$ (Andrews \& Williams 2005).

During stage I, both inner and outer embryos grow a hundredfold, reaching the critical masses $m_1 = 10.0\,M_{\oplus}$ and $m_2 = 11.0\,M_{\oplus}$, respectively; then the runaway mass growth is initiated. The mass increase during  stage I is approximated by an exponential law $m(t) = m_i \exp{(t/\tau)}$ ($i = 1, 2$),  where $m_i$ are the initial masses of the inner and outer planets  and the e-folding time is $\tau = 2.17\,\times\,10^5$\,years, for both planets. 
Although there is no widely accepted expression for the mass growth at this stage of planet formation, the exponential law seems to be a good approximation (Alibert et al. 2005, see Figures 5 and 8 in that paper); on the other hand, other authors considered different expressions, for instance, $sin^2 t$ (Lega et al. 2013). The type and the timescale of the planetary migration during stage I define the chosen initial values of the planet masses: they correspond to sufficiently small planets to experiment  an initial type I migration, but also to sufficiently large planets, in order to provide migration timescales on order of $8\,\times\,10^5$ years (see Armitage 2010).
\begin{table}
\centering
\caption{Boundaries of habitable zones (HZ) around our Sun and a star of mass $M_{\ast}=1.61\,M_\odot$ (Kopparapu et al. 2013).}
\begin{tabular}{l c c c }
\hline\hline
   Limit    &  Sun   & star   & ratio        \\
\hline\hline
         &            &               &\\
Inner HZ &      AU    &        AU     &\\
\hline
Runaway greenhouse limit & 0.97 & 2.20  & 2.27 \\
Recent Venus limit & 0.75 & 1.74  & 2.32 \\
         &            &               &\\
Outer HZ &     AU    &        AU      &\\
\hline
Maximum greenhouse limit & 1.70 & 3.80 & 2.24 \\
Early Mars limit  & 1.77 & 4.00   &  2.26   \\
\hline
\end{tabular}
\label{tableHZ}
\end{table}

We simulate the migration process during the first stage applying the semi-analytical models of Tanaka et al. (2002) and Tanaka $\&$ Ward (2004). The model approximates the decay and damping rates of a planet, with mass $m$, semimajor axis $a$, and eccentricity $e$, orbiting a star of mass $m_{\ast}$ and embedded in a laminar disk, as

\begin{equation}
\begin{array}{ccl}
\dot{a} &=& - \frac{2.7 + 1.1 q}{h^2} \frac{ m }{ m_*^{1.5}} \sqrt{{\mathcal G}} \Sigma_0 a^{(1.5-q)},   \\
 &  &  \\
\dot{e} &=&  \frac{e}{h^2} \frac{0.78}{(2.7+1.1 q)} \frac{\dot{a}}{a},
\label{eq1-2-2-1}
\end{array}
\end{equation}
where ${\mathcal G}$ is the gravitational constant and $h$, $q$, and $\Sigma_0$ are the scale height, the shape of the surface density profile, and the surface density of the disk at 1\,AU, respectively. It is worth mentioning that the model is valid only for low eccentricities $(e\,\leq\,0.05)$ and small planetary masses $(\leq\,13\,M_{\oplus})$.

In the case of a smooth density profile ($q\approx 0$), explicit expressions for the forces acting on the planetary masses can be found in Ogihara $\&$ Ida (2009) and Ogihara et al. (2010), while the applications of the model can be found in Giuppone et al. (2012). In our simulation, we assume a scale height $h = H/r = 0.05$ and a constant surface density profile $\Sigma (r) = \Sigma_0(r/1{\rm AU})^{-q}$ with $q = 0$ and $\Sigma_0 = 300\,g/cm^2$, which corresponds to two minimum mass solar nebula at 5\,AU (MMSN, Hayashi 1981). Assuming the median disk-to-star mass ratio of $\sim\,0.5\,\%$ (Andrews $\&$ Williams 2005), we consider the disk of the mass of 2 MMSN, which surrounds the HD\,200964 star with roughly $2\,M_{\odot}$ (MMSN corresponds to the Sun-type star with $1\,M_{\odot}$). From Eq. (\ref{eq1-2-2-1}), we estimate that, for the chosen disk parameters, the $e$-damping is approximately $100$ times faster than the $a$-decay.

The initial positions of the planets can be obtained using the definition of \textit{habitable zone} (HZ). For this, the boundaries of the HZ around our Sun and a MS star with the mass of  M=1.61\,$M_\odot$ were calculated following the approach given in Kopparapu et al. (2013), with the effective temperature and the luminosity of the star adopted at $7500$\,K and $6.4\,L_\odot$ (Inglis 2003). The results obtained are shown in Table \ref{tableHZ}. The comparison between the same limits around the Sun and the star shown in the last column allows us to obtain an averaged scaling factor equal to 2.27. Using this factor and the current distances of Jupiter and Saturn from the Sun (5.2\,AU and 9.5\,AU), we obtain the `analogous' initial distances of the planets in the HD\,200964 system. The values obtained are $a_1 = 13$\,AU and $a_2 = 22$\,AU, yielding the initial positions beyond the $2/1$ MMR, at $n_1/n_2\,\sim\,2.2$ (or $n_2/n_1\,\sim\,0.45$, see Figure \ref{fig1}).

The planetary orbits are assumed to be initially quasi-circular ($e_1 = e_2 = 0.02$) and co-planar. This choice is based on current migration models (e.g., Eq. \ref{eq1-2-2-1}) and hydrodynamical simulations which indicate that the damping of planetary eccentricities and inclinations during migration should be very rapid, with characteristic timescales much shorter than the timescale of the migration process itself. The eventual passages through resonances during migration may excite the eccentricities and/or inclinations, which are rapidly damped again as soon as the planetary pair leaves the resonance. However, when the system is captured in a mean-motion resonance, its dynamics is changed radically: the system will evolve now along a family of resonant stationary solutions and its eccentricities monotonously increase (Ferraz-Mello et al. 2003). The vertical stability of this evolution guarantees no excitation of planetary inclinations, which generally takes place at high eccentricities (for details, see Voyatzis et al. 2014). For this reason, we use the planar model to simulate the capture and evolution inside the 4/3 MMR. Finally, the initial values of the angular elements of both planets are chosen equal to zero.

\begin{figure}
\begin{center}
  \includegraphics[width=0.99\columnwidth]{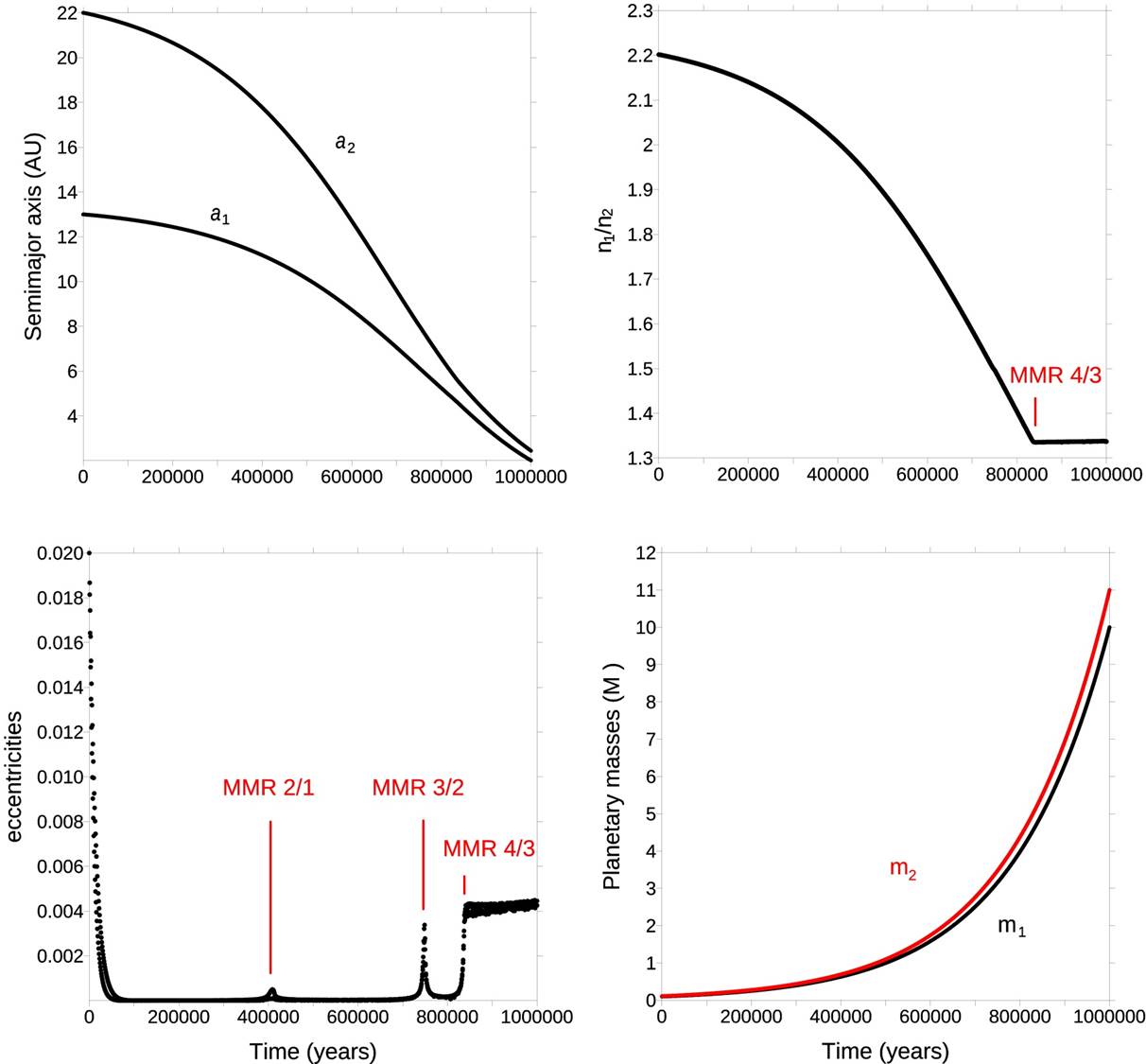}
\caption{Dynamical evolution of the planetary embryos in the protoplanetary disk. The mass increase of the planets is shown in the bottom-right panel. The orbital decay is shown in the top-left panel. The top-right panel shows the mean-motion ratio as a function of time. The evolution of the eccentricity is shown in the bottom-left panel, the crossing of the 2/1 and 3/2 MMRs are indicated. Finally, the capture in the 4:3 MMR occurs at $\sim 0.8$ Myr. }

\label{fig2}
\end{center}
\end{figure}

The described set of the disk and orbital parameters is referred hereafter to as `standard configuration' or run $no.$ 0; it is summarized in Table \ref{table1-2-2-2} (first row). The evolution of the corresponding system, during stage I of formation, is shown in Figure \ref{fig2}.

The exponential mass growth of the inner and outer planets to the critical values is shown by  the black and red curves, respectively, in the bottom-right panel in Figure \ref{fig2}. The top-left panel shows the decay of the planetary semimajor axes. The smooth evolution of the semimajor axes, even during the passages through the 2/1 and 3/2 MMRs, is noticeable, also in the top-right panel where the evolution of the ratio of the mean motions of the planets is shown. We note that, at $\sim\,8\,\times\,10^5$\,yr, the system approaches and is then trapped into the 4/3 MMR.

The evolution of the planet eccentricities is shown in the bottom-left panel in Figure \ref{fig2}. The passages through the 2/1 and 3/2 MMRs at $\sim\,4\,\times\,10^5$ and $\sim\,7.5\,\times\,10^5$\,years, respectively, provoke visible excitations of both eccentricities, which, because of the strong damping provoked by interactions with the disk, drop rapidly to zero as soon as the system leaves the resonances.  When the planets cross the 2/1 and 3/2 MMRs, their masses are still small, $\sim\,1.2\,M_{\oplus}$ and $\sim\,4\,M_{\oplus}$, respectively, and no resonant capture is observed. Approaching the 4/3 MMR at approximately $8\,\times\,10^5$\,years, the two planets grow to $\sim\,6.5\,M_{\oplus}$. The planets become sufficiently massive to mutually excite their eccentricities and remain in this state despite undergoing significant damping. This means that the pair of the planets is captured inside the 4/3 MMR and the evolution of the ratio of the planet mean motions shown in Figure \ref{fig2}\,\emph{top-right} confirm this fact.

The first stage of planet formation ceases after $ 10^6$\,years, with two planets locked in the 4/3 MMR. The masses of the inner and outer planets have increased to $10.0\,M_{\oplus}$ and $11.0\,M_{\oplus}$ ($\sim\,0.03\,M_J$), respectively. These masses correspond to the critical value for which the mass of the envelope is nearly equal to the mass of the planetary core (Alibert et al. 2005; Armitage 2010) and runaway growth of the planets is initiated (Pollack et al. 1996; Rice $\&$ Armitage 2003; Ikoma et al. 2000).


\subsection{Probability of the trapping in the 4/3 MMR during stage I}\label{sect-2-4}

To analyze the probability of the capture inside the 4/3 MMR, we tested different sets of the disk parameters and the initial planet configurations. Some of these sets are shown in Table \ref{table1-2-2-2}, where the first row (run $no.$ 0) corresponds to the standard configuration and the following rows show one-by-one (systematic) changes of the parameter of this configuration. The last two columns show the final configurations of the system at the end of stage I of the evolution. We have detected three possible final configurations: the trapping in the 4/3 MMR, the capture in the 3/2 MMR, and the non-resonant configuration caused by the diverging evolution of the planet orbits.
\begin{table*}
\begin{center}
\caption{Simulations performed with different orbital and disk parameters (see Sect. \ref{sect-2-4} for more details).} \label{table1-2-2-2}
\resizebox{14cm}{!} {
\begin{tabular}{ccccccccccc}
\hline
\hline
 & \multicolumn{3}{c}{ } & \multicolumn{5}{c}{  } & \multicolumn{2}{c}{  } \\
 run $no.$ & \multicolumn{3}{c}{Disk parameters} & \multicolumn{5}{c}{Initial planet  conditions }  & \multicolumn{2}{c}{Results} \\ [1.0ex]
\hline
 & & & & & & & & & & \\
  & q & h & $\Sigma_0$ & $a_1$ & $a_2$ & $n_1/n_2$  & $m_1$ & $m_2$ & $n_1/n_2 $ & $a_1$  \\[0.4ex]
    &   &   & MMSN & AU & AU &   & $M_{\oplus}$ & $M_{\oplus}$ & & AU \\[1.0ex]
\hline
 & & & & & & & & & & \\
  0 & 0.00 & 0.05 & 2 & 13 & 22 & 2.2  & 0.10 & 0.11 & $ 4/3$ & 2.00 \\[2.0ex]
  1 & 0.00 & 0.05 & 2 & 13 & 22 & 2.2  & 0.10 & 0.10 & $ 3/2$ & 2.04 \\
  2 & 0.00 & 0.05 & 2 & 13 & 22 & 2.2  & 0.11 & 0.10 & $ 3/2$ & 1.85 \\
  3 & 0.00 & 0.05 & 2 & 13 & 24 & 2.5  & 0.10 & 0.11 & $ 4/3$ & 2.02 \\
  4 & 0.00 & 0.05 & 2 & 10 & 17 & 2.2  & 0.10 & 0.11 & $ 3/2$ & 1.73 \\[1.0ex]
  5 & 0.25 & 0.05 & 2 & 13 & 22 & 2.2  & 0.10 & 0.11 & $ 3/2$ & 3.00 \\
  6 & 0.50 & 0.05 & 2 & 13 & 22 & 2.2  & 0.10 & 0.11 & 1.93 & 4.44 \\
  7 & 1.50 & 0.05 & 2 & 13 & 22 & 2.2  & 0.10 & 0.11 & 2.30 & 8.90 \\[1.0ex]
  8 & 0.00 & 0.07 & 2 & 13 & 22 & 2.2  & 0.10 & 0.11 & $ 3/2$ & 3.70 \\
  9 & 0.00 & 0.05 & 4 & 13 & 22 & 2.2  & 0.10 & 0.11 & $ 4/3$ & 0.74 \\[1.0ex]
\hline
\end{tabular}
}
\end{center}
\end{table*}

We have found that the probability of capture in the 4/3 MMR during type I migration is not sensitive to the variations of the initial eccentricities and angular orbital elements of the planets because the non-zero initial eccentricities are rapidly damped to zero-values as a result of disk-planet interactions and, consequently, have no effect on the capture into a mean-motion resonance. In addition, for nearly circular orbits, the orbital angles are circulating and can take any value in the range of $0-360^o$. Similar results were observed in the case of the 3/2 MMR (Correa-Otto et al. 2013).

On the contrary, the probability is sensitive to the variations of the masses of the planets and the initial semimajor axes of their orbits can be seen from the results of the runs from $1-4$ in Table \ref{table1-2-2-2}.  Indeed, if we assume that $m_2 \leq m_1$, the planets are captured inside the 3/2 MMR (runs $no.$ 1 and $no.$ 2). This result may be easily understood by analyzing Eq. (\ref{eq1-2-2-1}). In this case, $\stackrel{.}{a_2}\,<\,\stackrel{.}{a_1}$ (i.e. the decay of the outer planet is slower than the decay of the inner one), what provokes the slower convergence between the planetary orbits, not sufficient to cross the 3/2 MMR toward the 4/3 MMR. According to Eq. (\ref{eq1-2-2-1}), the convergence rate also depends on the initial scale of the system: it is faster for the more distant from the central star planet pairs, what is favorable for capture in the 4/3 MMR (compare the final configurations of the run $no.$ 3 and the run $no.$ 4).

The choice of the appropriate physical parameters of the disk seems to be essential for the capture inside the 4/3 MMR (runs $no.$ 5 - $no.$ 9 in Table \ref{table1-2-2-2}), especially of the shape of the surface density profile $q$. For the non-zero values of the parameter $q$, the probability of the capture inside the 4/3 MMR is low and the planet trapping does not occur in our few simulations (runs $no.$ 5 - $no.$ 7).  In fact, according to  Eq. (\ref{eq1-2-2-1}), $\dot{a}\,\sim\,a^{(1.5-q)}$ and the increasing values of $q$ will decrease the rate of convergence of two orbits. For low $q$-values (run $no.$ 5), the convergence rate is not sufficient for the non-capture passage through the 3/2 MMR; for higher values (run $no.$ 6), the convergence is so slow that the system does not even approximate the 3/2 MMR; and, finally, for $q = 1.5$ (run $no.$ 7), two orbits diverge increasing their mutual distance during the orbital decay. These results are in agreement with the results of investigations of the capture probability inside the first order resonances as a function of the rate of the orbital convergence shown in Mustill $\&$ Wyatt (2011).

Finally, the last two runs  ($no.$ 8 and $no.$ 9) investigate the probability of capture inside the 4/3 MMR as a function of the height $h$ and the surface density $\Sigma$ of the disk. For a higher disk (run $no.$ 8, with $h = 0.07$), the convergent velocity of the outer planet is not sufficient to cross the $3/2$ MMR, while for a more massive disk (run $no.$ 9), the convergence velocity allows the system to cross the 3/2 MMR and reach the 4/3 MMR, but in this case the orbital decays of the planets are very rapid and the final location of the planet pair is very close to the central star.

We conclude that the capture inside the 4/3 MMR demands a sufficiently fast convergence of the planetary orbits and, consequently, very specific disk and planet parameters are needed to trap the planets in this resonance during the first stage of formation.  The convergence rate is defined by many parameters, which must satisfy the following conditions i) the outer planet must be more massive than the inner one; ii) the planet pair must be located initially sufficiently far away from the central star; iii) the protoplanetary disk must be thin ($h\,\leq\,0.05$) and flat $(q\,<\,0.25)$; and iv) the mass of the disk must be $\sim$ 2\,MMSN. It is worth noting that the capture in the 4/3 MMR is possible for higher values of the surface density $\Sigma_0$ (e.g., run $no.$ 9 in Table \ref{table1-2-2-2}); however, in this case, the current distances of the planets from the central star strongly restrict the possible $\Sigma_0$--values.


\subsection{Stage II: runaway mass growth and type II migration}\label{sect-2-5}

\begin{figure}
\begin{center}
  \includegraphics[width=0.99\columnwidth]{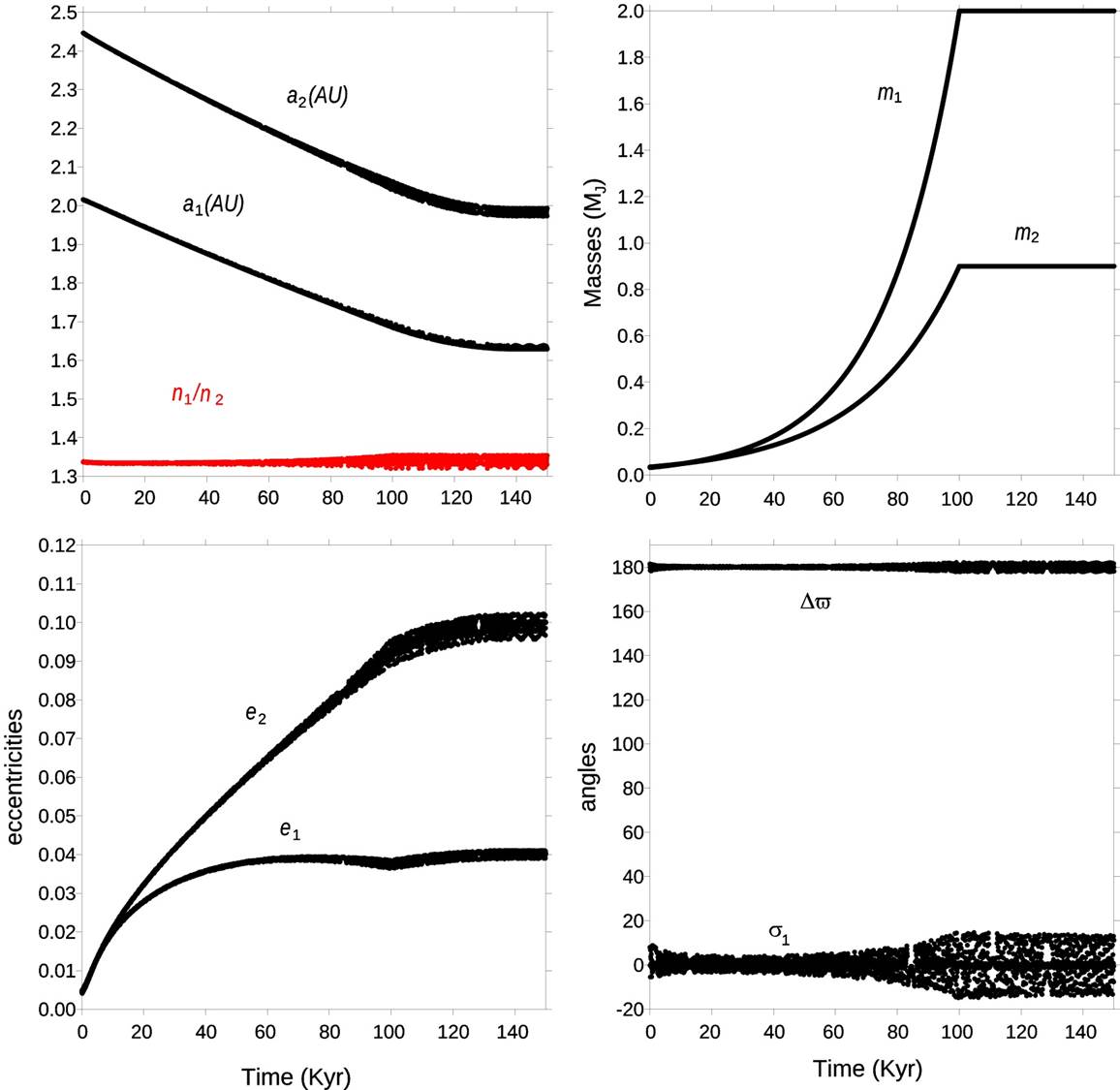}
\caption{Dynamical evolution of the planets in stage II of evolution. The mass increase of the planets is shown  the top-right panel. The orbital decay of the semimajor axes and the mean-motion ratio on the top-left panel. The bottom-left panel shows the eccentricity as a function of time, while the evolution of the resonant angle $\sigma_1$ and  the secular angle $\Delta \varpi$ is shown on the bottom-right panel. }
\label{fig3}
\end{center}
\end{figure}

The second stage of the planet formation begins when the envelope mass of the planet equates with its core mass; as a consequence, the accretion rate is accelerated dramatically and the runaway growth is initiated (Pollack et al. 1996; Ikoma et al. 2000; Armitage 2010). We assume that, for a star with mass $1.61\,M_{\odot}$, this phase lasts on average approximately $1.5\,\times\,10^5$\,years (Alibert et al. 2005; Armitage 2010; Hasegawa $\&$ Pudritz 2012). During this stage the planetary mass growth is approximated by an exponential law $m(t) = m_i \exp{(t/\tau_i)}$, where $m_i$ are the initial masses of the planets and $\tau_i$ are the e-folding times of the planet growth ($i = 1, 2$). The values of $m_i$ come from the end of stage I and are $10.0\,M_{\oplus}$ and $11.0\,M_{\oplus}$, for the inner and outer planets, respectively. The values of $\tau_i$ are chosen to be $2.4 \times 10^4$\,years and $3 \times 10^4$\,years, for the inner and outer planets, respectively, such that the planets reach their current masses ($2\,M_J$ and $0.9\,M_J$) after $10^5$\,years. Figure \ref{fig3}\,\emph{top-right} shows the exponential accretion of the planet masses during the second phase. The growth process is stopped after $10^5$ years, and the final values of the planetary masses are compatible with the best-fit solutions of Jonhson et al. (2011).

During the planet growth, the structure of the disk in the neighborhood of the planet is modified and gaps begin to form. From this stage onward we assume a type II of migration of the two planets which are already locked near the 4/3 MMR resonance. To simulate this migration, we applied a Stokes-like non-conservative force (Beaug\'e et al. 2006). The effects of the Stokes force on the evolution of the semimajor axis and the eccentricity are decay and damping, respectively, given by
\begin{equation}
a(t)= a_0\,exp{(-t/\tau_a)}  \\
{\rm and }\\
e(t)= e_0\,exp{(-t/\tau_e)} , \\
\end{equation}

where $a_0$ and $e_0$ are the initial values of the semimajor axis and the eccentricity, and $\tau_a$ and $\tau_e$ are the e-folding times of the corresponding orbital elements (Beaug\'e $\&$ Ferraz-Mello, 1993 ; Gomes, 1995). The planet configurations at the end of stage I of the evolution were used as input of the second stage.

The loss of the orbital energy during the migration of the planets toward the central star defines  the values of the e-folding times $\tau_a$. They are typically obtained through adjustment between the migration duration and the final locations of the planets, which are roughly $a_1\,\sim\,1.6$\,AU and $a_2\,\sim\,2$\,AU (Table \ref{table1-2}). If we assume that the second stage lasted $ 10^5$ years, we obtain $\tau_{a_1}=6.7 \times 10^5$ and $\tau_{a_2}=4.5 \times 10^5$\,years.

The choice of the e-folding times of the eccentricity damping is more complicated because the exchange of the angular momentum between the planets and the disk is very sensitive to the disk properties and the processes involved. In the simplified simulations based on the Stokes-like forces, it is common to introduce the factor $K=\tau_a/\tau_e$, which is the ratio of the e-folding times of the orbital decay and the eccentricity damping, and use it as a free parameter (Lee $\&$ Peale 2002; Beaug\'e et al. 2006; Kley et al. 2005).  The typical values of the factor K lie in the range from 1 to 100 (Lee $\&$ Peale 2002). We tested several values of K from this interval and we have found that, for $K = 1$, the configuration which the system acquires at the end of stage II,  matches the best-fit configuration obtained in Jonhson et al. (2011) and our best-fit solution (see Table \ref{table1-2}). The planetary masses and the orbital elements of the planets corresponding to this final configuration are given in Table \ref{table1-2-2-3}.
\begin{table}
\caption{Orbital parameters of the simulated main-sequence HD\,200964 system at the end of formation. }
\label{table1-2-2-3}
\centering
\begin{tabular}{l l l l r r}
\hline\hline
 Object &  $M$   & $a(AU)$   & $e$  & $\lambda$ (deg) & $\varpi$ (deg)        \\
\hline
   HD\,200964    & 1.61 $M_{\odot}$  &           &         &          &          \\
   HD\,200964 b  & 2.0 $M_J$              &  1.63     & 0.04    &   70      & 178      \\
   HD\,200964 c  & 0.9 $M_J$              &  2.00     & 0.10    &  104      &   0      \\
\hline
\end{tabular}
\end{table}

The time evolution of the simulated system during the second stage of formation obtained for K = 1 is shown in Figure \ref{fig3}. The migration process characterized by decay of the planetary semimajor axes is shown in the top-left panel: the timespan of the decay, defined by the value of $\tau_{a_1}$ and $\tau_{a_2}$, is extended to $\sim\,1.4 \times 10^5$\,years.
We can see that the magnitudes of the semimajor axes at the end of formation are similar to those obtained in Johnson et al. (2011) and of our best-fit solution (see Table \ref{table1-2}).

The behavior of the eccentricities and the characteristic angles $\sigma_1$ and $\Delta \varpi$ ($\sigma_1 - \sigma_2 $, see Eq. \ref{eq1-3-1}) during the second stage of migration is shown in Figure \ref{fig3}\,\emph{bottom-left}. As the migration is slow, the evolution of the eccentricities (bottom-left panel) closely follows the family of stationary solutions of the conservative 4/3 resonant problem, the ACR family (Beaug\'e et al. 2003, Michtchenko et al. 2006). The different rates of the mass accretion, when the inner planet grows more rapidly, are responsible for the excitation and the continuous increase of the eccentricity of the less massive outer planet (see Correa-Otto et al. 2013, for details). During the whole second stage the system remains inside the resonance, what is confirmed by the evolution of the characteristic angles $\sigma _1$ and $\Delta \varpi$ (Figure \ref{fig3}\,\emph{bottom-right}) .

During the last $5\,\times\,10^4$\,years of the process, the gas disk dissipates, while the pair of the already formed planets trapped in the 4/3 MMR (see Figure \ref{fig3}) continues to decay slowly (since the characteristic frequencies of the system evolution are, at least, one order higher, this process is still adiabatic). The disk is dissipates completely after $1.4\,\times\,10^5$\,years and the planets begin to evolve under only the central force and their mutual gravitational perturbations (the last $10^4$\,years in our simulation).

The total time of formation in our scenario is $1.15\,\times\,10^6$\,years. This timespan is compatible with the lifetime of protoplanetary disks, which is estimated to lie between 1-10\,Myr for a solar-mass and similar stars (Armitage 2010; Haisch et al. 2001).


\subsection{Post-main-sequence evolution of the HD\,200964 system} \label{sect-2-6}

The star HD\,200964 is currently a sub-giant K0 IV star, thus, the long-term stability of the planetary system in the presence of the post-main-sequence evolution of the central star should be analyzed. Following the study done in Voyatzis et al. (2013) and Voyatzis \& Hadjidemetriou (2006), we consider an exponential law for the loss of the mass of the HD\,200964 star during its post-main-sequence evolution

\begin{equation}
M(t)=M_0\, exp{(-t/\tau)},
\label{eq:Voyatzis}
\end{equation}

where $M_0 = 1.61\,M_{\odot}$ is the MS mass of the HD\,200964 star assumed in this work. The choice of the e-folding time $\tau$ is ambiguous since the duration of the sub-giant stage of the star is unknown. However, the value of $\tau$ may be constrained knowing that the sub-giant stage must be much shorter than the MS age of the star. If the total age of the star is assumed to be $3 \times 10^9$\,years, the timespan of $10^6$\,years seems to be a reasonable estimation of the sub-giant stage of the system. Indeed, this time is only 0.1\,\% of the age of the star, but it is much longer than the secular variations of the system (on the order of $10^2$\,years), such that the orbital changes are sufficiently slow to be considered as an adiabatic process. The e-folding time is then obtained as $\tau=10^7$\,years. If the current mass of the HD\,200964 star is $1.57\,M_{\odot}$ (Mortier et al. 2013), then the total loss of the MS mass after $10^6$\,years was $\sim\,3\,\%$.

In a similar study on the variable-mass two-body problem (Veras et al. 2011), the stellar mass decrease is approximated by $M(t) = M_0 - \alpha t $, with the constant rate $\alpha$.  In that paper, the authors introduce a dimensionless parameter, the `mass-loss index',  which is defined as
\begin{equation}
\Psi = \frac{\alpha}{n \mu},
\label{eq:Veras}
\end{equation}
with $n$ being the mean motion of a planet and $\mu$ being the sum of the planetary and stellar masses. They show that the planetary motion will be stable only when $\Psi \ll 1$ (adiabatic regime). Applying a simple fitting procedure, we obtain that, for the adopted parameters, the exponential mass decrease (Eq. \ref{eq:Voyatzis}) is approximated to the constant mass-loss decrease (Eq. \ref{eq:Veras}), when $\alpha \sim 4 \times 10^{-8}\,M_\odot\,yr^{-1}$. For this $\alpha$--value, $\Psi \ll 1$, and so the masses and mean motions of both planets, it is expected that the loss of the mass of the central star during its post-main-sequence stage will not raise instabilities in the  HD\,200964 planetary system.

\begin{figure}
\begin{center}
  \includegraphics[width=0.99\columnwidth]{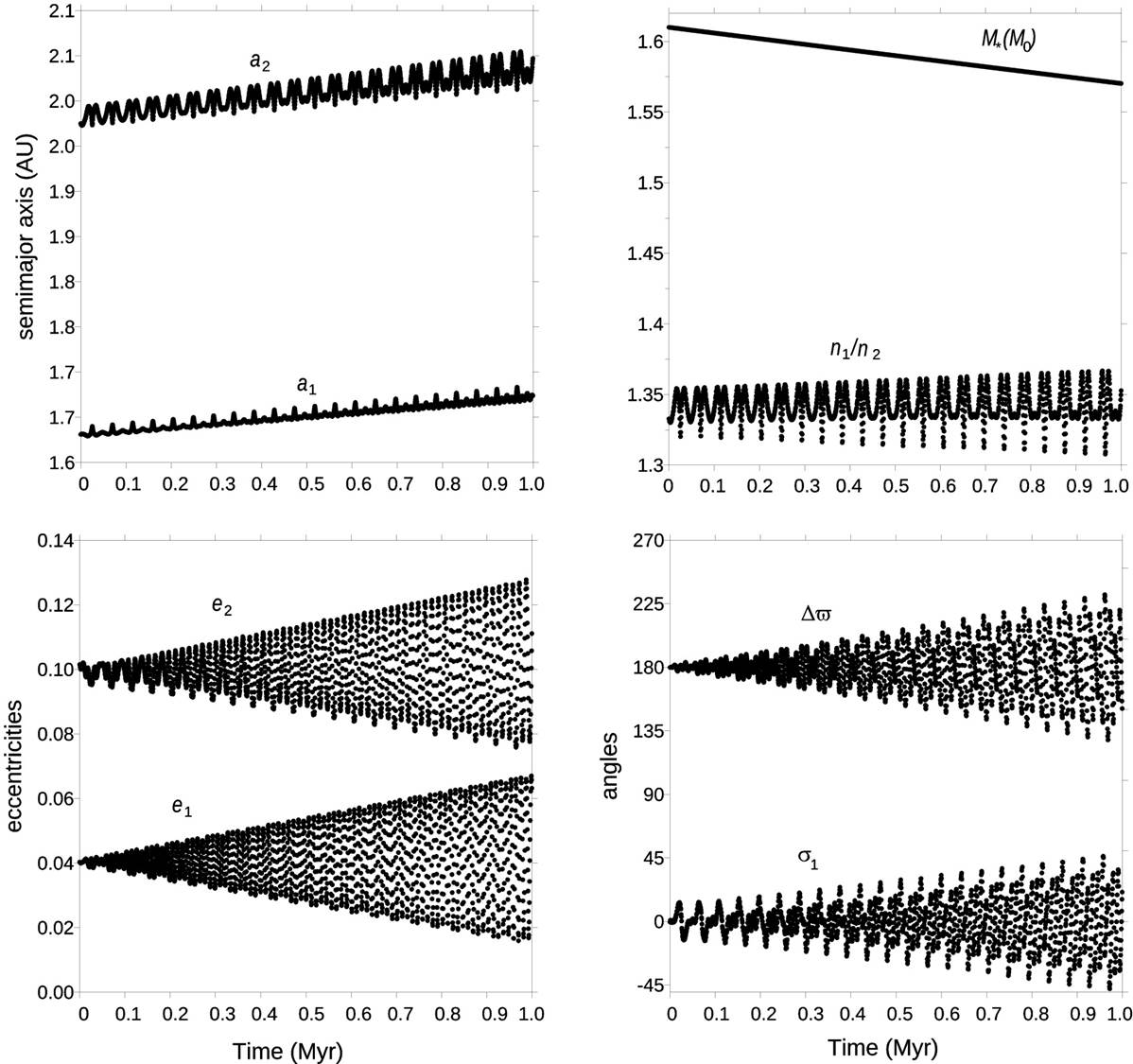}
\caption{Dynamical evolution of the planets in the presence of the post-main-sequence evolution of the star. The time evolution of the semimajor axes of the planets is shown on the top-left panel. The mass decrease of the star and the mean-motion ratio are shown on the top-right panel. The bottom-left panel shows the eccentricity as a function of time. The evolution of the characteristic angles $\sigma_1$ and $\Delta \varpi$ is shown in the bottom-right panel. }
\label{fig4}
\end{center}
\end{figure}

The evolution of the planetary orbits during the post-main-sequence evolution of the HD\,200964 star is shown in Figure \ref{fig4}. The top panels show the time evolution of the semimajor axes (top-left) and the mean-motion ratio, together with the stellar mass decrease (top-right). We can observe that the planets move slowly away from the central star, but their mutual resonant configuration remains unchanged.

The time evolution of the eccentricities and the characteristics angles is shown in the bottom panels in Figure \ref{fig4}. We can observe a continuous increase in the amplitudes of oscillations of all orbital elements. This behaviour indicates possible destabilization of the resonant configuration, which can occur when the amplitude of the resonant angle $\sigma_1$ (Eq. \ref{eq1-3-1}) reaches high magnitudes ($\sim 100^o$). However, according to results obtained in Veras et al. (2013) and Voyatzis et al. (2013), this does not occur in the adiabatic regime of the small mass loss ($\sim 3 \%$).

The dynamical stability of the HD\,200964 system obtained in our simulation (Table \ref{table1-2}, last column) was tested and confirmed applying the stochasticity indicator MEGNO (Cincotta $\&$ Sim\'o 2000); we have found that, for the simulated system,  $< Y > = 2$ in approximately $3000$\,years. The stability of the system was also tested through the direct numerical integration of the exact equations of motion over 10$^9$\,yr.

\begin{figure}
\begin{center}
  \includegraphics[width=0.99\columnwidth]{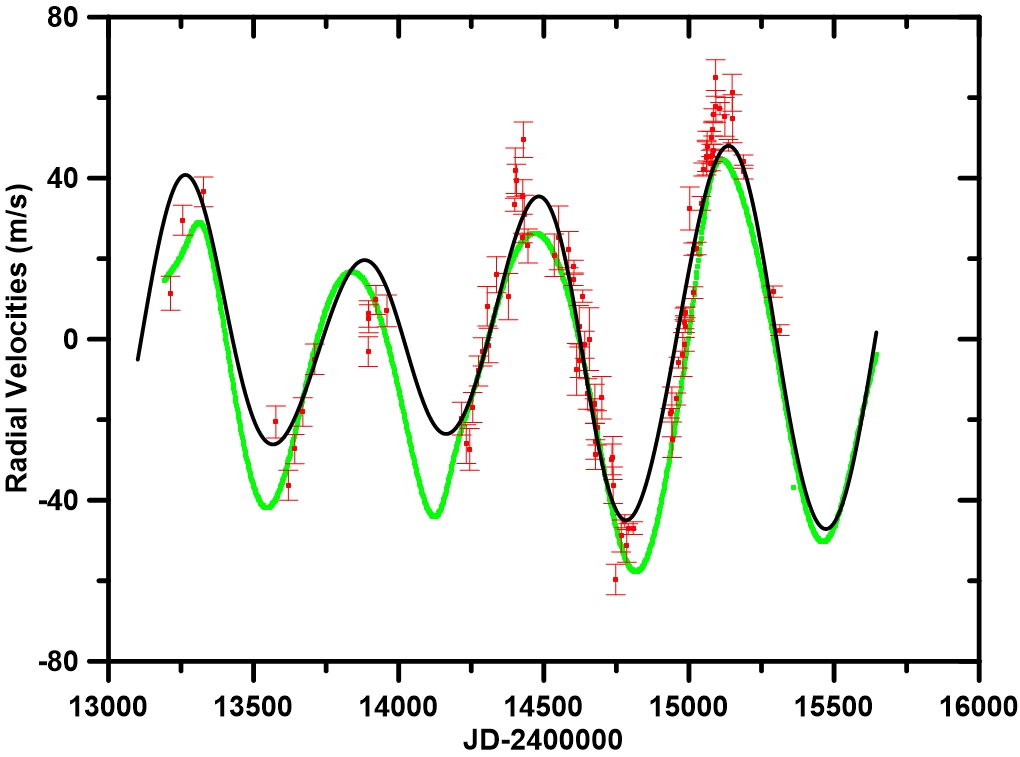}
\caption{Comparison between the radial velocities obtained for the simulated HD\,200964 system (black line) and our best-fit solution (green line) with the RV measurements (red dots). }
\label{fig5}
\end{center}
\end{figure}

Finally, the radial velocity of the star calculated for the simulated HD\,200964 system is shown in Figure \ref{fig5} by a black curve and can be compared with the Jonhson et al. (2011) observational RV data (red dots). We can see that our result is in good agreement with the observational data, with the $\sqrt{\chi_\nu^2}$ - value equal to 3.7 (we stress the fact that no effort was made to minimize the $\sqrt{\chi_\nu^2}$ - value by fine-tuning the parameters of the problem). In addition, this value of $\sqrt{\chi_\nu^2}$ is close to that resulting from the best-fit of the observations to an unconstrained Keplerian motion (green curve) obtained in Sect. \ref{sect-4}.
%


\section{Dynamical stability of the 4/3 MMR}\label{sect-3}

\subsection{Jonhson et al. (2011) orbital best-fit configuration}\label{sect-3-1}

The dynamical stability of the HD\,200964 planetary system was first discussed in the discovery paper by Jonhson et al. (2011); the orbital parameters of the nominal best-fit obtained there are presented in Table \ref{table1-2}. The stability of this configuration was tested over 10\,Myr and the involvement in the 4/3 mean-motion resonance has been suggested. Using those orbital elements, Wittenmyer et al. (2012) have mapped the neighborhood of the nominal best-fit configuration of the planets and have observed that the planetary motion can be regular over 100\,Myr only if the planets are currently trapped in the mutual 4/3 MMR.

\begin{figure}
\begin{center}
  \includegraphics[width=0.99\columnwidth]{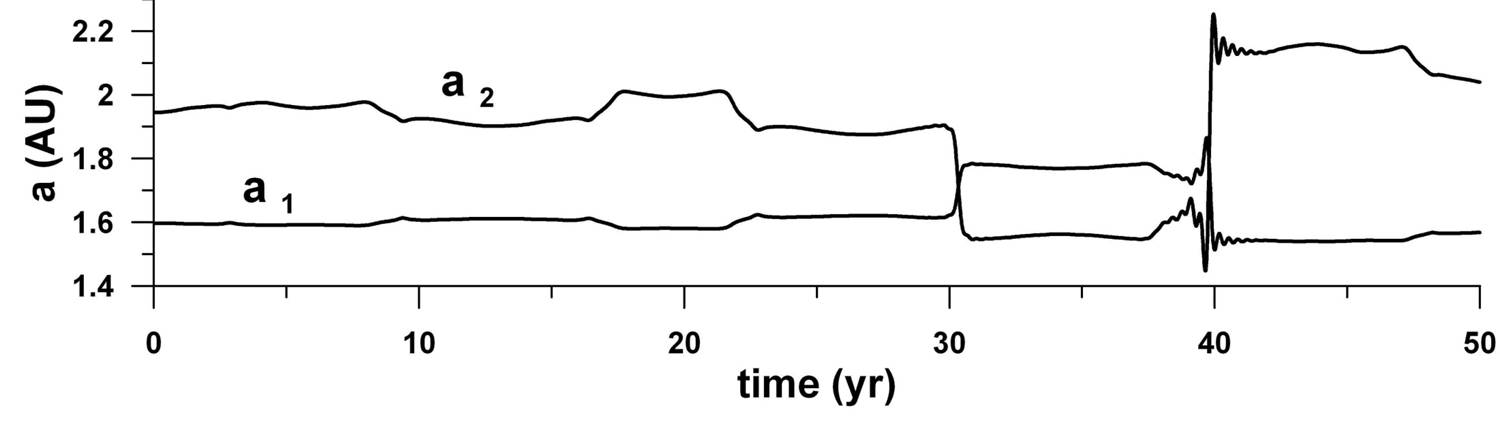}
\caption{Time evolution of the semimajor axes of the planets b and c corresponding to the nominal Jonhson et al. (2011) best-fit configuration.}
\label{fig6}
\end{center}
\end{figure}

We have also tested the dynamics of the planets around HD\,200964, using the stellar and planetary masses and the orbital elements from Jonhson et al. (2011), and have found that it is highly chaotic, with two planets colliding within a few tens of years, as shown in Figure \ref{fig6}. Since the authors have not specified the reference frame in which the orbits were obtained, we have tested several possible systems, such as heliocentric, Poincare, canonical, and Jacobian systems, and have obtained similar results for all of them.

It is known that the stability of resonant configurations is very sensitive to the values of the critical angles, characteristic of the mean-motion resonance. For the 4/3 MMR, these angles are defined as

\begin{equation}
\begin{array}{lll}
\sigma_1\,\,=\,\,4\lambda_2-3\lambda_1-\varpi_1&=&4M_2-3M_1+4\Delta \varpi, \\
\sigma_2\,\,=\,\,4\lambda_2-3\lambda_1-\varpi_2&=&4M_2-3M_1+3\Delta \varpi,
\label{eq1-3-1}
\end{array}
\end{equation}

where $\lambda_i$, $M_i$, and $\varpi_i$ are mean longitudes, mean anomalies, and longitudes of pericenter of the planets, and $\Delta \varpi=\varpi_2-\varpi_1$ (index 1 denotes the inner planet and index 2 denotes the outer planet). The behavior of the critical angles defines the location of the system with respect to the resonance: when one of the angles is librating, the system is said to be inside the resonance. Which one of the  angles is librating depends on the ratio of the planetary masses and will be specified in the following.

\begin{figure}
\begin{center}
  \includegraphics[width=0.99\columnwidth]{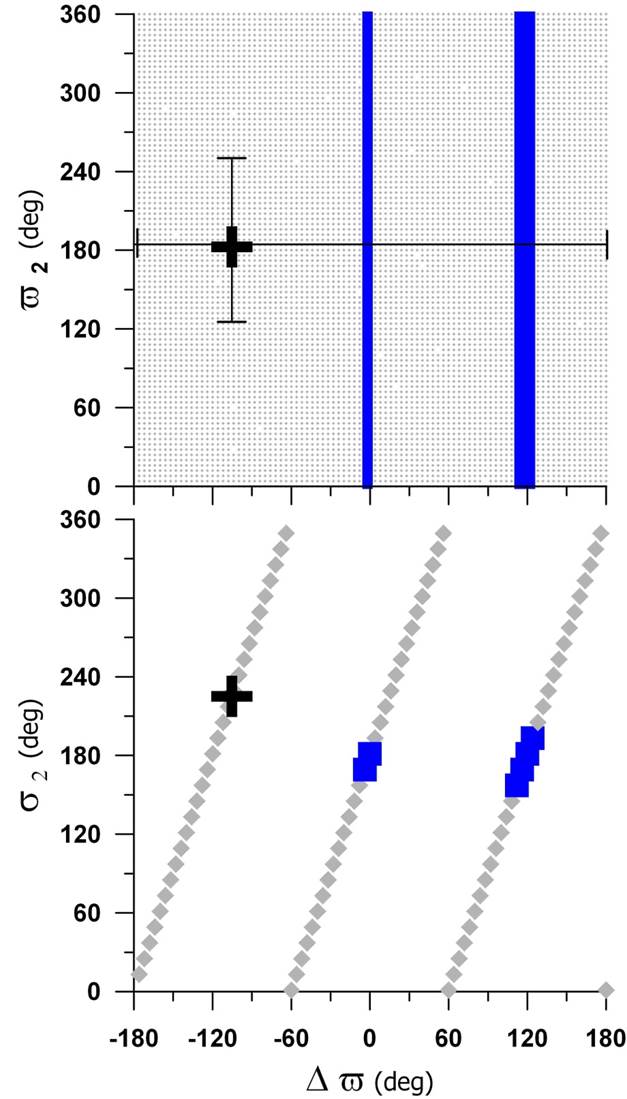}
\caption{Top panel: Stability map on the $(\Delta \varpi,\varpi_2)$--plane of initial conditions (all other elements are fixed at those given by the best-fit of Jonhson et al. 2011): blue symbols show configurations stable over 7\,Myr, while gray symbols show orbits unstable over this time interval. Location of the best-fit configuration from Jonhson et al. (2011) is shown by a cross, together with the uncertainty bars. Bottom panel: Transformation of the top panel to the $(\Delta \varpi, \sigma_2)$--plane.}
\label{fig7}
\end{center}
\end{figure}

The longitudes of the pericenter of the planetary orbits, $\varpi_1$ and $\varpi_2$, are known to be poorly determined quantities in the fitting procedures (see uncertainties in Table \ref{table1-2}). Keeping unchanged all other orbital elements from Jonhson et al. (2011), we tested the stability for possible values of these angles, in the range from 0 to 360$^o$.  Figure \ref{fig7}\,\emph{top} shows the results on the $(\Delta \varpi, \varpi_2)$--plane of initial conditions, where $\Delta \varpi = \sigma_1 - \sigma_2$ is the secular angle of the problem. The systems surviving over 7\,Myr are shown by blue symbols in Figure \ref{fig7}\,\emph{top}, while those with ejected/collided planets within this time interval, are shown by gray symbols. The location of the nominal best-fit solution of Jonhson et al. (2011) on this plane is shown by a cross, together with the uncertainty bars. We note that the uncertainty in the $\Delta\varpi$--determination covers the whole range from $0$ to $360^o$. We can note in Figure \ref{fig7}\,\emph{top} that the system corresponding to the best-fit is located inside the sea of strongly unstable motions, what confirms the result of the simulation shown in Figure \ref{fig6}.

The $(\Delta \varpi,\varpi_2)$--plane can be transformed in the $(\Delta \varpi, \sigma_2)$--plane of the characteristic angles of the 4/3 MMR using Eq. (\ref{eq1-3-1}) with the values of the mean anomalies  of the planets, $M_1$ and $M_2$, fixed at $176^o.0$ and $87^o.3$, respectively,  according to the Jonhson et al. (2011) fit. The map obtained is shown in the bottom panel in Figure \ref{fig7}. It shows that stable motions (over, at least, 7\,Myr) are tightly confined in the domains around $\sigma_2=180^o$. It is worth noting that, in the nominal best-fit configuration of Jonhson et al. (2011), $\sigma_2 = 225^o$; the significant deviation from $180^o$ can explain the unstable behaviour shown in Figure \ref{fig6}.  The confinement of $\sigma_2$ around 180$^o$, for stable solutions, shows that it is a truly resonant librating angle of the HD\,200964 system; as shown in  (Michtchenko et al. 2008\,b), the resonant systems with the outer planet smaller than the inner one are generally in the `external resonance', which is characterized by the libration of $\sigma_2$ and circulation/oscillation of $\Delta \varpi$ (and $\sigma_1$). The condition $\sigma_2 = 180^o$ can be used as a constrain in the determination of the orbits of the HD\,200964 system.

%

\subsection{Dynamical maps of the 4/3 MMR} \label{sect-3-2}

The stability of the planetary motion inside the 4/3 MMR is very sensitive to the mutual orbital configuration of the planets. Indeed, in this resonance, the orbits are so close that, even for small eccentricities, they may cross. The resonant configuration is the only one which may prevent the close approaches and collision between the giant planets, but this configuration requires very specific mutual locations of the planets. The analysis of the stable resonant configurations is generally done using the Hamiltonian models of the resonance (e.g., Michtchenko et al. 2006). Here we only briefly introduce some basic properties of the resonant motion.

\begin{figure}
\begin{center}
  \includegraphics[width=0.99\columnwidth]{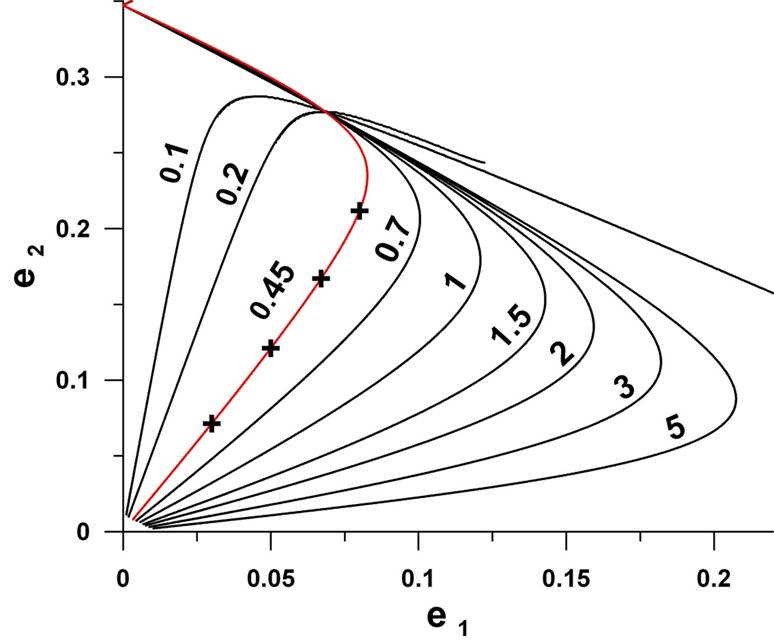}
\caption{The 4/3 resonance stable ACRs parameterized by the mass ratio $m_2/m_1$. The family with the mass ratio 0.45 corresponding to the simulated HD\,200964 system (Table \ref{table1-2}, last column) is shown by a red curve. The crosses are ACR whose neighborhood was analyzed and whose stability maps are shown in Figure \ref{fig9}.}
\label{fig8}
\end{center}
\end{figure}

The resonant motion is stable when the system is evolving in the close vicinity of the stable stationary solution of the averaged resonant problem. These solutions are often referred to as apsidal corotation resonances (ACRs). To obtain ACRs of the 4/3 MMR, we employ the geometrical method described in Michtchenko et al. (2006); they are shown on the ($e_1$,$e_2$) eccentricities plane in Figure \ref{fig8}, in the form of families parameterized by the ratio $m_2/m_1$ of the planetary masses. The family parameterized by the mass ratio $0.45$ corresponding to the solution simulated  in Sect. \ref{sect-2} is shown by a red curve. The scaling parameter which defines the distances of the planets to the central star, depending only on the semimajor axes (see Michtchenko et al. 2008), is the same for all solutions, while the total angular momentum is specific for each ACR shown in Figure \ref{fig8}.

The resonant (critical) angles given in Eq. (\ref{eq1-3-1}) define the mutual orientation of the orbits and the positions of the planets on them. All solutions shown in Figure \ref{fig8} have the critical angles $\sigma_1$ and $\sigma_2 $ fixed at $0$ and $180^o$, respectively, the last value corresponding to the detected stable motions shown in Figure \ref{fig7}. Thus, in the stable stationary configurations, the planetary orbits are anti-aligned and, in conjunction (when the mutual distance is minimal), the inner planet is at the pericenter of its orbit, while the outer planet is at apocenter. This geometry is mostly favorable for the stable evolution of the 4/3 resonance system.

Knowing the ACRs of the 4/3 MMR, we can study the planetary dynamics around them. We have chosen four ACRs to analyze their vicinity; they all belong to the family parameterized by $m_2/m_1 = 0.45$ corresponding to the planet pair whose formation was simulated in Sect. \ref{sect-2} (see Table \ref{table1-2}, last column), and are shown by cross symbols in Figure \ref{fig8}. The analysis is done in the form of a stability map on the representative ($n_2/n_1,e_2)$--plane, where $n_1$ and $n_2$ are the planetary mean motions. In its construction, the domain around one ACR is covered with a rectangular grid of initial conditions, with spacings $\Delta (n_2/n_1)=0.001$ and $\Delta e_2 =0.001$. Each ACR under study corresponds to two constants of motion, the angular momentum and the scaling parameter (Michtchenko et al. 2008\,a). Fixing these constants at the values of the corresponding ACR, we calculate the values of the semimajor axis and the eccentricity of the inner planet at each node of the grid. In the case when the solution exists, the test system is numerically integrated over 3\,Myr, starting at  $\sigma_1$  fixed at 0 and  $\Delta \varpi$ fixed at either 0 (positive values on the e2-axis in Figure \ref{fig8}) or $180^o$ (negative values on the e2-axis). It should be noted that, although the chosen timespan of 3\,Myr is very short compared to the lifetime of the central star, it is sufficiently long to detect the main dynamical mechanisms that can destabilize the planetary motion. Thus, we do not need to perform very long (but yet not conclusive) simulations, similar to those ($\sim$\,100\,Myr) done in Wittenmyer et al. (2012), to confirm the stability of the system over the star's lifetime.

\begin{figure*}
\begin{center}
  \includegraphics[width=0.99\textwidth]{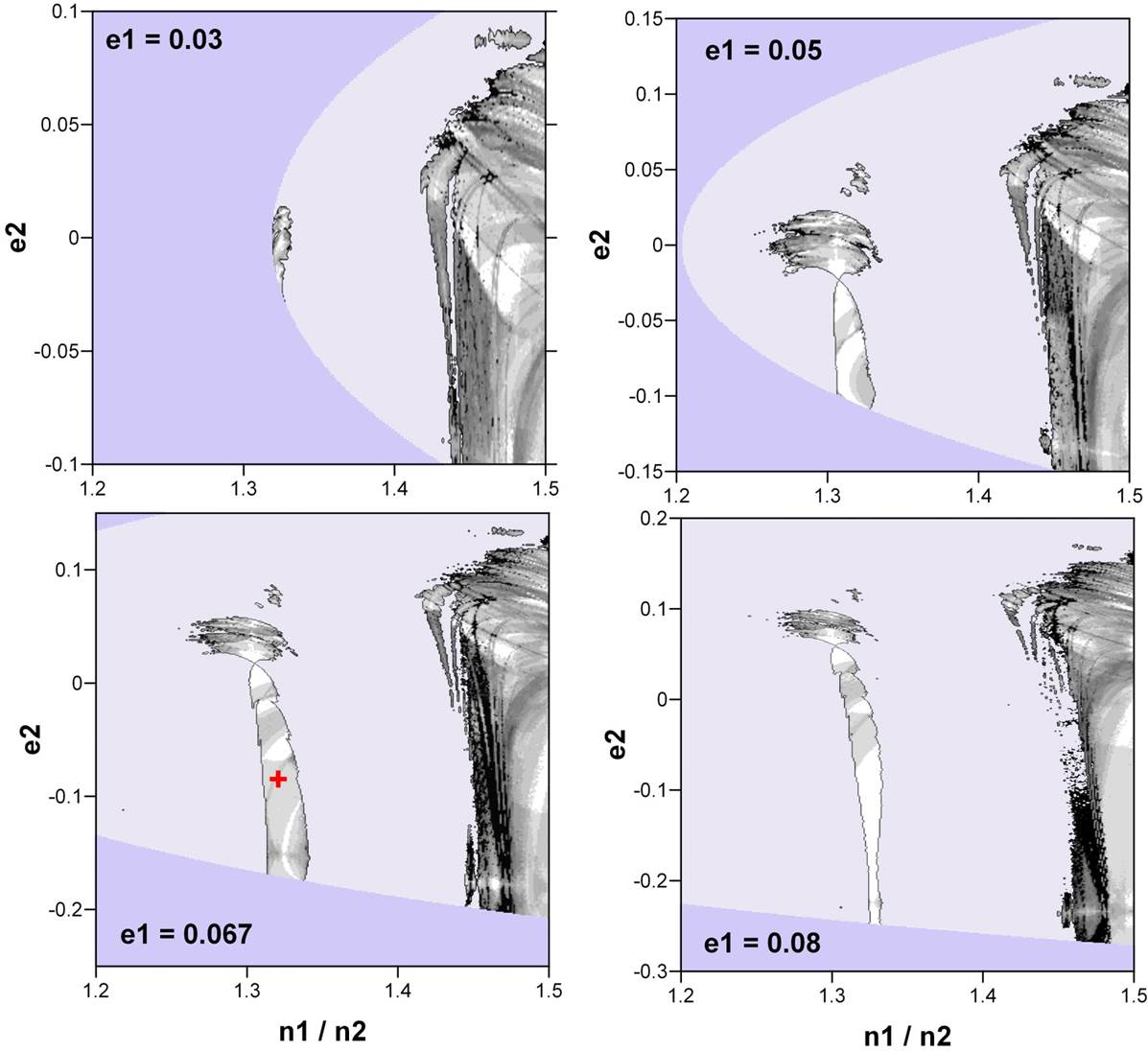}
\caption{Dynamical maps of the domain around the stable symmetric ACRs with $e_1 = 0.03$ (top left) and $0.05$ (top right); and, $e_1 = 0.067$ (bottom left) and $0.08$ (bottom right) with $m_2/m_1 = 0.45$. The light tones correspond to stable motions, while the increasingly dark tones correspond to increasingly chaotic motions. The dark-blue regions correspond to initial conditions for which no solutions exist for the fixed values of the angular momentum and the scaling parameter. Finally, the domains of initial conditions leading to the collision/ejection of the planets within 3\,Myr are light-blue colored. The initial value of $\sigma_1$ is fixed at 0 and the initial value of $\Delta \varpi$ is fixed at 0 (positive values on the e2-axis) or $180^o$ (negative values on the e2-axis). The red cross in the bottom left plane shows the location of the HD\,200964 system simulated in Sect. \ref{sect-2}. }
\label{fig9}
\end{center}
\end{figure*}

The stability maps obtained for four different ACRs are shown in Figure \ref{fig9}. Each plane is characterized by the values of the inner planet eccentricity $e_1$ of the corresponding ACR, that can be easily identified in Figure \ref{fig8}. The stability of one orbit was tested using the SAM method, briefly described in Sect. \ref{sect-1} (for details, see Michtchenko et al. 2002, Ferraz-Mello et al. 2005, Appendix A.3). 

The 4/3 MMR is located around $n_1/n_2 = 1.33$, while the close 3/2 MMR is located around $n_1/n_2 = 1.5$. Both resonances are strong first-order order resonances; however, as seen in Figure \ref{fig9},  the region of the 4/3 MMR is very narrow, while the domain of the 3/2 MMR is large. These distinct manifestations of the resonances can be explained by the actions of the overlapping high-order mean-motion resonances which occur between the very massive and very close planets.
\begin{figure}
\begin{center}
  \includegraphics[width=0.99\columnwidth]{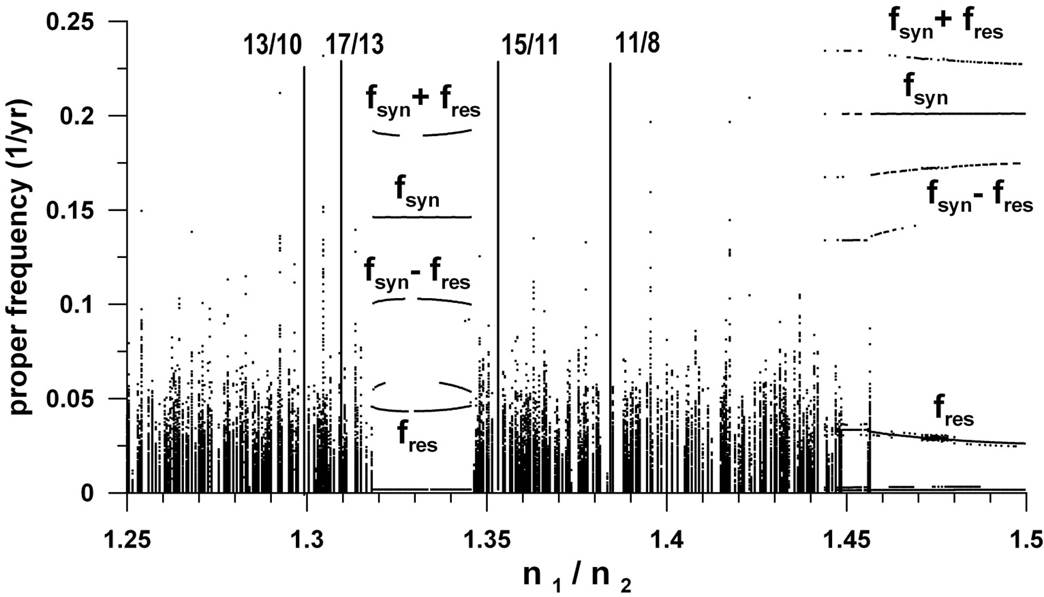}
\caption{Dynamical power spectrum showing the proper frequencies of the tested systems as functions of the mean-motion ratio. The smooth evolution of the frequencies is characteristic of regular motion and occurs around $n_1/n_2=1.33$ (4/3 MMR) and $n_1/n_2=1.5$ (3/2 MMR). The erratic scattering of the points in the other domains is characteristic of the chaotic motion. The nominal location of several high-order MMR is shown.}
\label{fig10}
\end{center}
\end{figure}
%


\subsection{Dynamical power spectrum of the 4/3 MMR} \label{sect-3-3}

To illustrate how the high-order MMRs work to destabilize the planetary motions in the 4/3 MMR, we present in Figure \ref{fig10} the dynamical power spectrum of the HD\,200964 system (for details, see Michtchenko et al. 2002, Ferraz-Mello et al. 2005, Appendix A4). The spectrum shows the evolution of the main frequencies of the systems as functions of the mean-motion ratio $n_1/n_2$.  The spectrum was calculated for the systems from the neighborhood of the ACR whose dynamical map is shown in the bottom-left plane in Figure \ref{fig9}, along the horizontal line on this plane with co-ordinate $e_2=0.098$. The smooth evolution of the frequencies is characteristic of regular motion and occurs in the region of the 4/3 MMR around $n_1/n_2=1.33$ and in the region of the 3/2 MMR close to $n_1/n_2=1.5$. The erratic spreading of the frequency values characterizes the strongly chaotic behaviour of the system.

The proper frequencies are identified as the synodic frequency (angular velocity of the synodic longitude), $f_{\rm syn}$, the resonant frequency of the libration of the critical angle $\sigma_2$,  $f_{\rm res}$, and the very low (close to zero) frequency of the secular angle $\Delta\varpi$. The simple linear combinations of the $f_{\rm syn}$ and $f_{\rm res}$ frequencies can also be easily identified; of these, only the ($f_{\rm syn}\pm f_{\rm res}$)--components are tagged  in Figure \ref{fig10}.

Figure \ref{fig10} shows that, inside the 3/2 MMR, the resonant frequency $f_{res}$ and the synodic frequency $f_{syn}$ are well separated. According to the 3$^o$ Kepler law, when the planets become closer, the synodic frequency decreases, in such a way, that $f_{syn}$ and $f_{res}$ become comparable and a beating between their harmonics may occur producing a number of high-order MMRs inside the 4/3 MMR. This phenomenon is known as `overlapping resonances' and its effects produce highly unstable motions of the planets (Wisdom 1980). Indeed, the boundaries of the regular motion inside the 4/3 MMR are clearly defined by the 11/8 and 15/11 MMRs at the right-hand side, and 17/13 and 13/10 MMRs at the left-hand side of the 4/3 MMR; the nominal positions of these MMRs are shown in Figure \ref{fig10}.

\begin{figure}
\begin{center}
  \includegraphics[width=0.99\columnwidth]{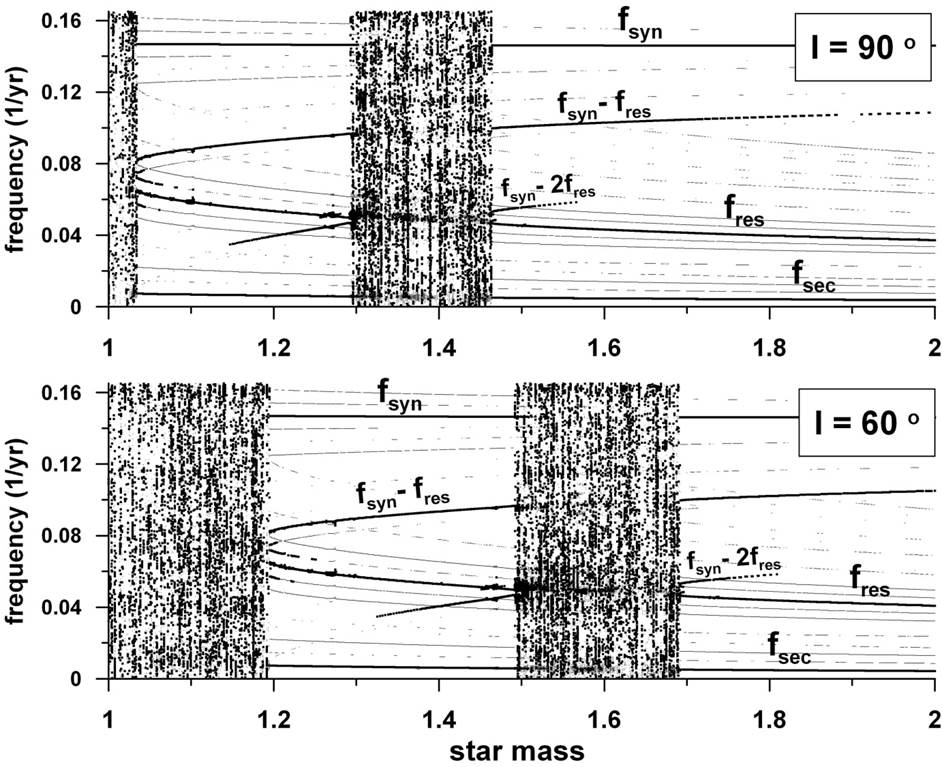}
\caption{Dynamical power spectra of the simulated solution, obtained with $m_1$sin\,$I$\,=\,2.0\,$M_J$ and $m_2$sin\,$I$\,=\,0.9\,$M_J$ ($I$ is the inclination of the planetary orbits to the sky plane), showing the proper frequencies of the system as functions of the mass of the central star. Top: spectrum obtained for sin\,$I$\,=\,90$^o$. Bottom: spectrum obtained for sin\,$I$\,=\,60$^o$. The smooth evolution of the proper frequencies  is characteristic of the regular motion, while the erratic scattering of the points is characteristic of the chaotic motion.}
\label{fig11}
\end{center}
\end{figure}

In what follows we show an example of how the above described dynamics allows us to constrain the parameters of the 4/3 resonant system, particularly, the individual masses of the planets affected by the sin\,$I$ indetermination ($I$ is the inclination of the orbits with respect to the sky plane). For this reason, we construct two dynamical power spectra around our simulated solution (Table \ref{table1-2}, last column), using the mass of the star as a parameter which varies from 1\,$M_\odot$ to 2\,$M_\odot$, while the planetary masses defined as $m_i\sin I$, $i=1,2$, are fixed. The spectra are shown in Figure \ref{fig11}. The top panel shows the spectrum obtained for $I=90^\circ$, which corresponds to the edge-on masses $m_1=2.0\,M_J$ and $m_2=0.9\,M_J$, while the bottom  panel shows the spectrum obtained for $I=60^\circ$, corresponding to $m_1=2.31\,M_J$ and $m_2=1.04\,M_J$.

Three proper frequencies of the system are observed in both graphs in Figure \ref{fig11}. The secular frequency $f_{sec}$ and its harmonics form a very-low-frequency band at $\sim\,0.003\,{\rm yr}^{-1}$. The 4/3 resonance frequency $f_{res}$ and its linear combinations with $f_{sec}$ form a band which decreases from  $\sim\,0.07\,{\rm yr}^{-1}$ to $\sim\,0.04\,{\rm yr}^{-1}$ with the increasing star mass. Finally, the short-period synodic frequency $f_{syn}$ located around $\sim\,0.15\,{\rm yr}^{-1}$ seems to be only slightly affected by the change of the star mass.

For $I=90^\circ$ (top graph), the chaotic behaviour raised by the overlapping resonances is observed in the two intervals of the star mass values: between $1.0\,M_\odot$ and $1.04\,M_\odot$ and between $1.30\,M_{\odot}$ and $1.47\,M_{\odot}$. The former interval corresponds to the condition when $f_{res}$ is nearly equal to $f_{syn}-f_{res}$, while the last interval corresponds  to the condition $f_{res}\,\sim\,f_{syn}-2\,f_{res}$, or, in terms of the MMRs, to the 7/5 and 11/8 ratios of the mean motions, respectively. For $I=60^\circ$ (bottom graph), the first interval is extended from $1.0\,M_\odot$ to $1.2\,M_\odot$, and the second interval lay between $1.50\,M_{\odot}$ and $1.69\,M_{\odot}$.

If we assume that the actual HD\,200964 system is edge-on (i.e., sin\,$I$=90$^o$), the plot in the top graph in Figure \ref{fig11} allows us to constrain the mass of the star and explain why the Jonhson et al. (2011) best-fit configuration, obtained for $M_{\ast} = 1.44\,M_{\odot}$, shows highly chaotic behavior (see Figure \ref{fig6}). However, the improved estimation of the stellar mass, given in Mortier et al. (2013) as $M_{\ast} = 1.57\,M_{\odot}$, yields the  stable motion of the system, according to Figure \ref{fig10}. On the other hand, if the stellar mass is well determined, we can put constrains on the inclination (that is, on the planetary masses). Indeed, according to the bottom graph in Figure \ref{fig11}, for the star mass equal to $1.57\,M_{\odot}$,  the inclination must be higher than $I$\,=\,60$^o$; that is, planetary masses must be less than $2.31\,M_J$ and $1.04\,M_J$, for the inner and outer planets, respectively. It is worth mentioning that the estimated line-of-sight inclination ($I$) is for an isotropic distribution of extrasolar systems in the Milky Way.

We have thus shown that that the planetary motions in the HD\,200964 system occur in extreme conditions, with two very massive planets evolving under mutual perturbations on very close, and even crossing, orbits. In such systems, besides the 4/3 MMR, the short-period interactions (on the order of synodic period) also play a significant role on the stability of the motion due to appearance of the high-order MMRs which, overlapping with the 4/3 MMR, produce strong instabilities. We suggest that the configurations of the real planets in the HD\,200964 system, allowing very narrow stability margins, require a new analysis of the observation data, which will be done in the next section.


\section{Analysis of radial velocities.}\label{sect-4}

\subsection{Discussion of the observations}

The radial velocities of the planets orbiting around HD\,200964 indicate the presence of two Jupiter-like planets (minimum masses $\sim\,1.8$ and $1.2\,M_J$), in orbits whose periods are close to the 4/3 commensurability, which, even for the large observed semimajor axes (1.6 and 2.0\,AU) is immerse in a highly chaotic domain. The most favorable results obtained using Johnson's initially determined orbits and masses indicate a timelife shorter than $10^8$\,years (Wittenmyer et al. 2010).

In what follows we present a new discussion of these observations aiming at an exploration of the boundaries of the confidence interval of the given resonant solution and also at search of possible solutions not in the 4/3 resonance. For that, the existing observations were analyzed using a biased Monte Carlo (BMC) procedure associated with $\chi^2$ improvement criterion (see Ferraz-Mello et al. 2005; Tadeu dos Santos et al. 2012, Beaug\'e et al. 2012). This procedure uses a standard orbit improvement technique (actually a genetic algorithm followed by simplex interpolations) with starting values taken at random in a broad set of initial conditions. Each run is aborted after some improvement of the testing parameter ($\chi^2$) and the resulting solution is compared to the observations. The results obtained in this way are generally bad, and are discarded. Only solutions that can be considered as good fits are kept. The cut-off level is chosen so as to allow the construction of a representative sample of good-fit solutions. Each solution in this sample is an independent set of 12 orbital elements and parameters. The solution corresponding to the minimum $\chi^2$ is the best-fit solution. Since the actual minimum of $\chi^2$ is an important factor in the analysis, a real least-squares solution is also obtained.

\begin{table}
\caption{Best-fit solution and $1\sigma$ confidence domains}
\label{table:1}
\centering
\begin{tabular}{l | l l | l l }
\hline\hline
 Parameter    &  \multicolumn{2}{|c}{without jitter} & \multicolumn{2}{|c}{with $3$\,m/s jitter}   \\
              &         median  &   confidence        &     median   &   confidence   \\
              &         values  &    interval        &       values  &   interval     \\
\hline
$P_1$(d)  & $ 646.8$ & $\left[ 610;645 \right] $  & $627.8$ & $\left[ 598;655 \right]$ \\
$P_2$(d)  & $ 885.8$ & $\left[ 797;1064 \right] $ & $828.3$ & $\left[ 770;990 \right]$ \\
$P_1/P_2$  & $ 0.72$ & $\left[ 0.57;0.8 \right] $ & $0.76$ & $\left[ 0.58;0.8 \right]$ \\
$a_1$(AU) & $1.646$ & $\left[ 1.59;1.67 \right] $ & $1.62$ & $\left[ 1.57;1.68 \right]$ \\
$a_2$(AU) & $ 2.030$ & $\left[ 1.9;2.3 \right] $ & $1.94$ & $\left[ 1.85;2.07 \right]$ \\
$m_1 \sin{I} (M_J)$ & $1.820$ & $\left[ 1.56;2.0 \right] $ & $1.85$ & $\left[ 1.08;2.3 \right]$ \\
$m_2 \sin{I} (M_J)$ & $0.956$ & $\left[ 0.84;1.3 \right] $ & $1.27$ & $\left[ 0.94;1.6 \right]$ \\
$e_1$ & $0.091$ & $\left[ 0.004;0.17 \right] $ & $0.087$ & $\left[ 0.001;0.18 \right]$ \\
$e_2$ & $0.451$ & $\left[ 0.130;0.60 \right] $ & $0.079$ & $\left[ 0.001;0.45 \right]$ \\
$\omega_1$(deg) & $181$ & $\left[ 80;341 \right] $ & $240.0$ & $\left[ 200;300 \right]$ \\
$\omega_2$(deg) & $305$ & $\left[ 264;352 \right] $ & $279.6$ & $\left[ 200;460 \right]$ \\
$\lambda_1$(deg)& $354.0$ & $\left[ 310;365 \right]$ & $328.8$ & $\left[300;370 \right]$ \\
$\lambda_2$(deg)& $278.8$ & $\left[ 230;390 \right]$ & $239.9$ & $\left[210;300 \right]$ \\
$K_1$(m/s)     & $33.9$  & $\left[ 29;37 \right]$  & $24.5$ & $ \left[ 31;43  \right]$  \\
$K_2$(m/s)     & $17.8$  & $\left[ 15;23 \right]$  & $21.7$ & $ \left[25;27 \right]$  \\
\hline
 $w.r.m.s$.    &  \multicolumn{2}{|c}{$5.18$\,m/s}  &  \multicolumn{2}{|c}{$6.27$\,m/s} \\
\hline
$\sqrt{\chi_\nu^2}$ &  \multicolumn{2}{|c}{2.47}  &  \multicolumn{2}{|c}{1.56} \\
\hline
\end{tabular}
\end{table}

The first columns of results in Table \ref{table:1} show the best-fit solution. We also give the interval of confidence corresponding to the $1\sigma$ level, which is defined, for each variable, by values taken by that variable in the set of all samples with $\chi^2$ smaller than the limit corresponding to the $1 \sigma$ deviation as defined by the classical formula given by Press et al. (1986, Sect. 14.5). For a complete discussion of the operations done, see Beaug\'e et al. (2013; Sect. 2.4.).

\begin{figure}
\begin{center}
  \includegraphics[width=0.99\columnwidth]{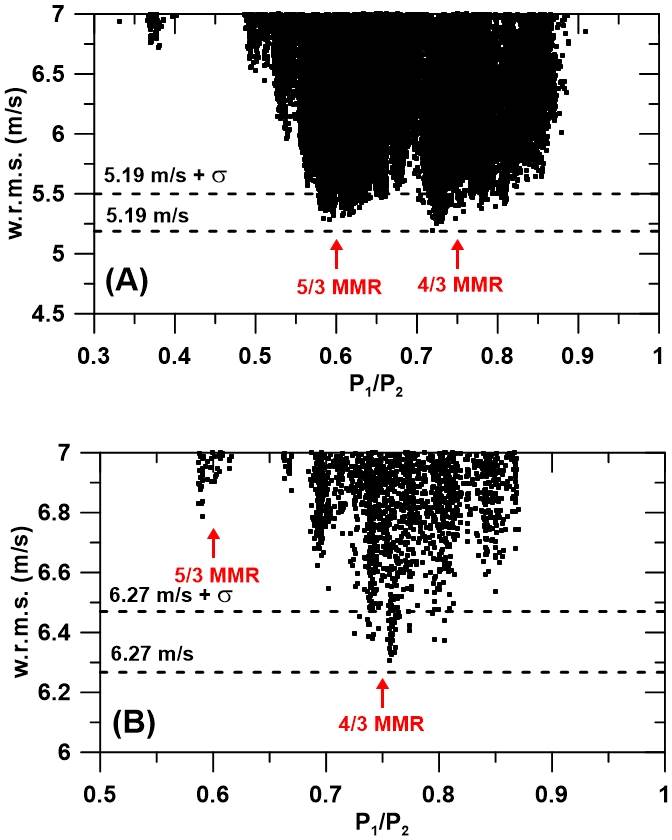}
\caption{Ratio of periods calculated by BMC procedure (see the text for details). In (A) we present the results without jitter, the horizontal dashed lines point to the best $w.r.m.s$ and $1\sigma$ domain. The red arrows mark the nominal values to 5/3 and 4/3 MMR. The panel (B) is the same as (A), but we insert a jitter of $3$\,m/s. }
\label{fig12}
\end{center}
\end{figure}

Surprisingly the sample constructed in this way showed not only the 4/3 resonant best-fit, but also one equivalent minimum at the 5/3 resonance. As seen in Figure \ref{fig12}, there are solutions in the neighborhood of the 5/3 resonance of quality equivalent to the best-fit solution (with weighted root mean square, $w.r.m.s.=5.25$\,m/s instead of the minimum $5.16$\,m/s).

The other striking point is the broadness of the sample of good solutions. For example, solutions in which the period $P_2$ lies between $797$ to $1064$\,days are found in the $1\sigma$ domain of confidence. The same broadness is seen in all variables. If we keep in mind that the meaning of the $1\sigma$ domain of confidence is that the probability that the actual solutions belong to it is roughly $70\,\%$, these results means that the observations do not allow us to determine one orbit or a bunch of orbits actually representing the system and, given the very high level of chaos present in this domain, the probability of getting on stable orbit from the observations is practically zero. 

We note that we have preferred to use the $w.r.m.s$ of the residuals instead of the $\chi^2$. These two quantities are related by the simple formula

\begin{equation}
\chi^2 = \frac{N-1}{N} \sum^{N}_{k=1} \frac{1}{\sigma_k^2} \left[ w.r.m.s \right] ^2,  \\
\end{equation}

where N is the number of observations. The $w.r.m.s$ was preferred since it can be interpreted independently of statistical hypotheses and also because it is less affected by a possible underestimation of the standard errors $\sigma_k$  of the individual observations.

It is necessary to mention that the values given in Table \ref{table:1} were obtained using a star of mass $1.44\,M_{\odot}$. Some recent determinations indicate a larger value, $1.57\,M_{\odot}$ (Mortier et al. 2013). If this larger value is adopted, the parameter $m \sin{I}$ given in Table \ref{table:1} must be increased in the same proportion. 
The orbital parameters with this larger stellar mass are illustrated in Table \ref{table1-2}, ant the curve of radial velocities is shown in Figure \ref{fig5} (green curve).

\subsection{Influence of the jitter.}

If no jitter is considered, the result obtained is strongly determined by the Keck observations. To assess the difference between the observations of the data from the two telescopes, it is enough to note that when the Keck observations are considered alone, the $w.r.m.s$ of the best-fit solution is only $1.8$\,m/s, instead of the $5.18$\,m/s of the full set. The Lick observations play an important role because they extend the observed interval in time far beyond the interval covered by the Keck observations, but at a price of a great deterioration in the $w.r.m.s$ of the results.

The introduction of one jitter changes the relative weight of the Keck and Lick observations, thus affecting the results. Johnson et al. used a $5$\,m/s jitter. However, the $w.r.m.s$ of the best-fit solution of the Keck observations ($1.8$\,m/s) shows that it is possible to assume a smaller jitter. We adopted the value $3$\,m/s. The results with this jitter are shown in the last column of Table \ref{table:1}. The plot of the confidence interval in $P_1/P_2$ (Figure \ref{fig12}) shows that the minimum of $\chi^2$ near the resonance 5/3 is now higher than the $1\sigma$ level. These solutions no longer belong to the $1\sigma$ domain of confidence and, in the light of the existing observations, can be discarded. However, future precise Keck observations of this system added to the existing set, may force a reconsideration of this solution.

\subsection{Influence of aliases}

Finally, we mention two technical problems related to these determinations. One of them is that the periods are almost symmetrical with respect to the Nyquist period of 2 years. This could be due to the influence of an incorrect consideration of one period and its aliases. Solutions with only one planet were tried but none of them was able to give a solution with a $w.r.m.s.$ better than $8.3$\,m/s (without jitter). The improvement of the results when two planets are considered seems real.

\begin{figure}
\begin{center}
  \includegraphics[width=0.99\columnwidth]{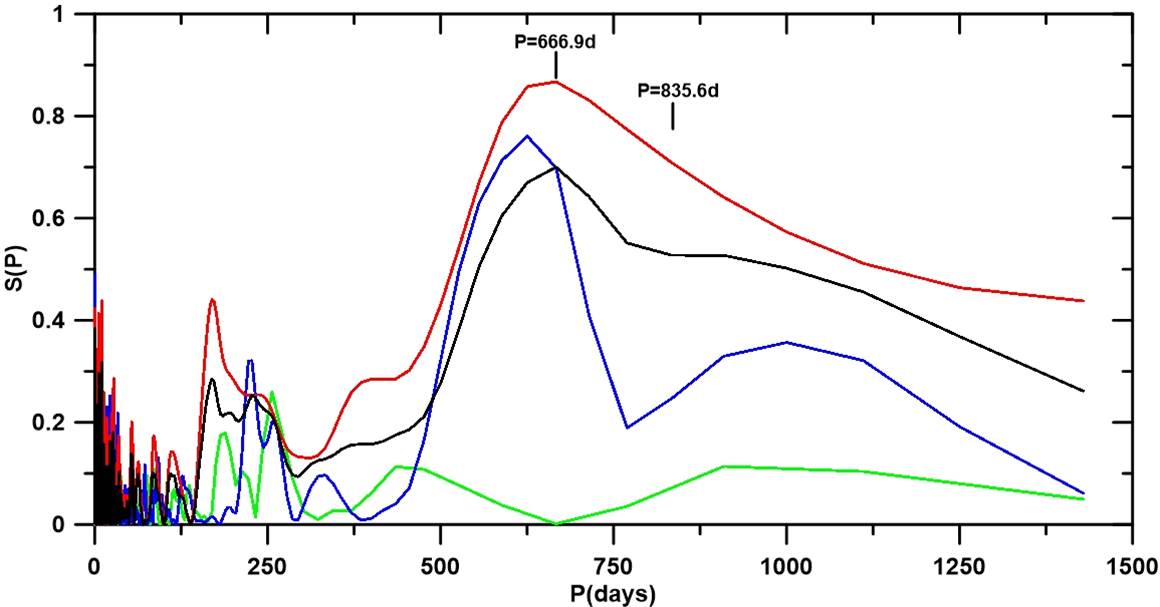}
\caption{Black curve is the DCDFT (date compensated discrete Fourier transform) normalized spectra of Radial Velocities given by Johnson et al. (2010). The red and blue curves are the spectrum of the Keck's and Lick's observations, respectively. In green curve we represent the spectra of the residuals after elimination of the period  $\sim$\,660\,days by one-body fit. We note the periods given by the best-fit of Johnson et al. (2010).}
\label{fig13}
\end{center}
\end{figure}

The other problem is the presence of several secondary peaks in the Fourier spectrum near the period 200\,days (see black curve in Figure \ref{fig13}). The Fourier spectrum (see Ferraz-Mello, 1981) of the residuals of the above two-planet determinations, however, does not show these peaks (see green and blue curves in Figure \ref{fig13}), indicating that those peaks result from beats of the two periods and the sampling periods.


\section{Summary }\label{sect-5}

In this paper, we have presented a complete dynamical and cosmogonical study of the HD\,200964 planetary system. The main topics in our analysis were i) a new determination of the orbits of the planets from the RV measurements, ii) the complete  comprehension of the planetary evolution in the vicinity of and inside the 4/3 MMR, and iii) the elaboration of a possible scenario of the formation for systems in the 4/3 MMR.

For the origin of the HD\,200964 system, we considered the scenario of formation of close planetary systems developed in Correa-Otto et al. (2013). This formation model was chosen after the detailed investigation of the planetary dynamics in the neighborhood of the 4/3 MMR. The study was done in form of dynamical maps (Michtchenko et al. 2002, Ferraz-Mello et al. 2005), which showed that the resonance capture would be possible only in the stage when the planets were still embryo-sized. Moreover, in our scenario we included the evolution of the star from the main sequence to the sub-giant branch.

With the star in the main sequence, we considered that the migration process of the planets began at the same time with planetary mass growth. The initial planetary masses were chosen to be on the order of planetary embryos and, at the end of the formation process, they achieved giant planet masses. In this way, the system formation scenario involved both type I and type II migration sets. The total timespan of simultaneous accretion and migration was assumed to have lasted $\sim\,10^6$\,years, which is compatible with the lifetimes observed for protoplanetary disks. We have found that the passage across the 2/1 and 3/2 MMRs, without capture inside them, is possible during the type I migration when planetary masses are small. However, the capture probability is highly dependent on the protoplanetary disk parameters. We have found that only a thin, vertically isothermal and laminar disk, with a nearly constant surface density profile, allows the system to cross the natural barriers created by the 2/1, 5/3, and 3/2 MMRs and to evolve towards the 4/3 MMR. Finally,  we considered the change of the mass of the star during its evolution from the main sequence to the sub-giant branch. We have obtained  that the stellar mass loss was $\sim$\,3\,\% and that the planetary motions were not destabilized.

The final orbital configuration found in the formation scenario (see Table \ref{table1-2}, last column) was used to calculate a radial velocity curve which was then compared to the existing observations. The result has a $\sqrt{\chi_\nu^2}$ value close to that resulting from the best-fit of the observations to an unconstrained Keplerian motion. Moreover, the planetary configuration obtained was tested in order to verify the dynamical stability of the system.  This was done in two ways: (i) we made numerical integrations of the exact equations  of motion over extended timespans, and (ii) we studied the general  dynamics of systems involved in the 4/3 resonance applying classical theories of mean-motion resonances (Beaug\'e et al. 2003, Ferraz-Mello et al. 2005, Michtchenko et al. 2006, Michtchenko et al. 2008\,a,b). In both cases, the long-period  stability of the simulated HD\,200964 planetary pair was confirmed.

The dynamical maps around the ACR solutions showed that the stability of the 4/3 MMR, differently from the 3/2 MMR, is restricted to a small region of the phase space  (see Figure \ref{fig9}). Moreover, according to Figure \ref{fig7}, the angular elements of the stable orbits are confined to narrow bands around $\Delta \varpi = 0$ and $120^o$, that, for given values of the mean anomalies ($M_1$ and $M_2$ fixed at $176.0^o$ and $87.3^o$, respectively, of the Johnson et al. 2011 best-fit solution) implies that the critical angle $\sigma_ 2$ is strictly confined to the domain around $180^o$. We have shown that the domains of stable motions inside the 4/3 MMR are reduced as a result of the influence of the short periods on the dynamics of the planetary pair. Indeed, for some initial configurations, the period of synodic angle beats with the resonant period, forming high-order mean-motion resonances and producing a strongly unstable motion (see Figure \ref{fig10}).

The phenomenon described above allowed us to understand the influence of stellar mass on chaotic behavior of the planets. According to Figure \ref{fig11}\,top, for $M_{\ast} = 1.44\,M_{\odot}$, we have the beating  $f_{syn}\,\sim\,3 f_{res}$ as source of instability, while for $M_{\ast} = 1.57\,M_{\odot}$ we observe a regular behavior of the planets. The stability of the solution was thus extremely sensitive to the stellar mass adopted; for this reason we were successful in getting the system formed with the stellar mass of $1.57$ and not $1.44\,M_{\odot}$.

In the course of this work, we reexamined the available radial velocities. This study was done, initially, in order to confirm the 4/3 MMR and$/$or find other possible configurations, different from the initially proposed one. Using the observations without any hypothesis, we have found, in addition to the 4/3 MMR (with $wrms = 5.2\,m/s$), another set of solutions in the 5/3 MMR ($wrms = 5.2\,m/s$). However, an analysis, taking stellar jitter into account allowed us to discard the 5/3 MMR solution.

The nominal values in our best solution do not differ from the Johnson et al. (2011) solutions, because they are also unstable in a short time. However, instead of looking just at the best-fit, we paid attention to the interval of possible solutions given in Table \ref{table:1}, and noted the fact that the stable regions found in Section 4 and the final conditions from the formation scenario are included in the set of possible solutions. By combining both, we find a set of statistically good and stable solutions, which are compatible with our scenario of formation.


\section*{Acknowledgments}

This work was supported by the S\~ao Paulo State Science Foundation, FAPESP, and the Brazilian National Research Council, CNPq. This work has made use of the facilities of the Computation Center of the University of S\~ao Paulo (LCCA-USP) and of the Laboratory of Astroinformatics (IAG/USP, NAT/Unicsul), whose purchase was made possible by the Brazilian agency FAPESP (grant 2009/54006-4) and the INCT-A. The authors are grateful to the anonymous referee for numerous suggestions and corrections to this paper.



\end{document}